\documentclass[fleqn,reqno]{amsart}

\usepackage{amscd,amssymb,amsmath,graphicx,verbatim}
\usepackage[pdftex]{hyperref}
\usepackage{textcomp}
\usepackage[OT1,T1]{fontenc}
\usepackage{mathabx}

\newtheorem{theorem}{Theorem}

\newtheorem{definition}{Definition}

\begin{document}

\title[Maximal quantum mechanical symmetry] {Maximal quantum mechanical symmetry: Projective representations
of the inhomogenous symplectic group}
\author{Stephen G. Low}
\date{\today}
\maketitle
\begin{abstract}

A symmetry in quantum mechanics is described by the projective representations
of a Lie symmetry group that transforms between physical quantum
states such that the square of the modulus of the states is invariant. The
Heisenberg commutation relations, that are fundamental to quantum
mechanics, must be valid in all of these physical states. This paper
shows that the maximal quantum symmetry group, whose projective
representations preserve the Heisenberg commutation relations in
this manner, is the inhomogeneous symplectic group. The projective
representations are equivalent to the unitary representations of
the central extension of the inhomogeneous symplectic group. This
centrally extended group is the semidirect product of the cover
of the symplectic group and the Weyl-Heisenberg group. Its unitary
irreducible representations are computed explicitly using the Mackey
representation theorems for semidirect product groups.
\end{abstract}
\section{Introduction}

The Heisenberg commutation relations are
\begin{equation}
\left[ {\widehat{P}}_{i},{\widehat{Q}}_{j}\right] =i \hbar  \delta _{i,j}\text{\boldmath
$1$},%
\label{MG: Heisenberg commutation relations}
\end{equation}

\noindent where $i,j,...=1,...,n$.\ \ The hermitian operators ${\widehat{P}}_{i}$
and ${\widehat{Q}}_{j}$ represent quantum mechanical momentum and position
observables acting on states $|\psi \rangle $ that are elements
of a Hilbert space $\text{\boldmath $\mathrm{H}$}$ for which $\text{\boldmath
$1$}$ is the unit operator.\ \ (We will use natural units in which
$\hbar =1$ throughout the paper.) These relations are fundamental
to quantum mechanics in its original formulation.\ \ 

Weyl \cite{weyl} established that these relations are the Hermitian
representation of the algebra of a Lie group $\mathcal{H}( n) $
that we now call the Weyl-Heisenberg group. The Weyl-Heisenberg
Lie group is a semidirect product \cite{folland} of two abelian
groups\footnote{In our notation for a semidirect product $\mathcal{G}\simeq
\mathcal{K}\otimes _{s}\mathcal{N}$, $\mathcal{N}$ is the normal
subgroup (see Definition 1 in Appendix A).\ \ Also, $\mathcal{A}\simeq
\mathcal{B}$ is the notation for a group isomorphism.}
\begin{equation}
\mathcal{H}( n) \simeq \mathcal{A}( n) \otimes _{s}\mathcal{A}(
n+1) ,%
\label{MG: WH definition}
\end{equation}

\noindent where $\mathcal{A}( m) $ is the abelian Lie group isomorphic
to the reals under addition, $\mathcal{A}( m) \simeq (\mathbb{R}^{m},+)$.\ \ Therefore,
it has an underlying manifold diffeomorphic to $\mathbb{R}^{2n+1}$
and is simply connected.\ \ In a global coordinate system $p,q\in
\mathbb{R}^{n}$, $\iota \in \mathbb{R}$, the group product and inverse
of the Weyl-Heisenberg group may be written
\begin{gather}
\Upsilon ( p^{\prime },q^{\prime },\iota ^{\prime }) \Upsilon (
p,q,\iota ) =\Upsilon ( p^{\prime }+p,q^{\prime }+q,\iota +\iota
^{\prime }+\frac{1}{2} \left( p^{\prime }\cdot q-q^{\prime }\cdot
p\right) ) ,%
\label{MG: Heisenberg group product}
\\{\Upsilon ( p,q,\iota ) }^{-1}=\Upsilon ( -p,-q,-\iota ) .%
\label{MG: Heisenberg group inverse}
\end{gather}

\noindent The identity element is $e=\Upsilon ( 0,0,0) $.\ \ Its
Lie algebra is given by
\begin{equation}
\left[ P_{i},Q_{i}\right] = \delta _{i,j}I,\ \ \left[ P_{i},I\right]
=0, \left[ Q_{i},I\right] =0.%
\label{MG: WH algebra}
\end{equation}

The faithful unitary irreducible representations $\xi $ of the Weyl-Heisenberg
group may be written as
\begin{equation}
\psi ^{\prime }( x) =\left( \xi ( \Upsilon ( p,q,\iota ) ) \psi
\right) \left( x\right) =e^{i \lambda ( \iota +x\cdot p-\frac{1}{2}p\cdot
q) }\psi ( x-q) %
\label{MG: Unitary representaton of WH}
\end{equation}

\noindent where $p,q,x\in \mathbb{R}^{n}$, $\iota \in \mathbb{R}$.
$\lambda \in \mathbb{R}\backslash \{0\}$ label the irreducible representations
and $\psi ( x) =\langle x|\psi \rangle \in {\text{\boldmath $\mathrm{H}$}}^{\xi
}\simeq {\text{\boldmath $L$}}^{2}( \mathbb{R}^{n},\mathbb{C}) $.\ \ We
label the Hilbert space with the unitary representation $\xi $ as
this Hilbert space, on which the unitary representation $\xi $ acts,
is\ \ determined by the unitary irreducible representation and is
not given {\itshape a priori}.

The Stone-von Neumann theorem \cite{Stone}, \cite{vonNeumann} establishes
that (6) defines the complete set of faithful irreducible representations
of the Weyl-Heisenberg group. This theorem is not constructive;
it does not give a prescription to obtain these representations
but only establishes that they are a complete set of faithful irreducible
representations. However, as the Weyl-Heisenberg group has the form
of the semidirect product given in (2), the unitary irreducible
representations (6) can also be directly calculated using the Mackey
theorems as these theorems are constructive. This is reviewed in
Section 3.1. 

The position and momentum operators in (1) are given by the faithful\footnote{There
are also degenerate representations corresponding to the homomorphism
$\pi :\mathcal{H}( n) \rightarrow \mathcal{A}( 2n) $ for which $\lambda
=0$ (See Appendix A, Theorem 4). These representations of the abelian
group are not discussed further here.} hermitian representation
$\xi ^{\prime }$ of the Weyl-Heisenberg algebra. (The prime designates
the lift of the unitary representation $\xi $ of the group to the
algebra, $\xi ^{\prime }=T_{e}\xi $.)
\begin{equation}
 {\widehat{P}}_{i}=\xi ^{\prime }( P_{i}) , {\widehat{P}}_{i}=\xi ^{\prime
}( P_{i}) , \widehat{I}=\xi ^{\prime }( I)  . 
\end{equation}

These operators also act on the Hilbert space ${\text{\boldmath
$\mathrm{H}$}}^{\xi }\simeq {\text{\boldmath $L$}}^{2}( \mathbb{R}^{n},\mathbb{C})
$.\ \ As the representation is a homomorphism, its lift preserves
the Lie bracket,
\begin{equation}
\left. \upsilon ^{\prime }( \left[ P_{i},Q_{i}\right] ) =\left[
\upsilon ^{\prime }( P_{i}) ,\upsilon ^{\prime }( Q_{i}) \right]
\right) =\ \ i\ \ \delta _{i,j}\upsilon ^{\prime }( I) =i\ \ \lambda
\delta _{i,j}\text{\boldmath $1$}.%
\label{MG: irrep Weyl-Heisenberg algebra}
\end{equation}

\noindent The $i$ appears simply because we are using hermitian
rather than anti-hermitian operators\footnote{In some neighborhood,
the group element $g$ is given in terms of an element\ \ $X$ of
the algebra by $g=e^{X}$. Then for a unitary representation $\upsilon
$ of the group, the representation $\upsilon ^{\prime }$of the algebra
is Hermitian (rather than non-Hermitian only if we insert an $i$,
$\upsilon ( g) =e^{i \upsilon ^{\prime }( X) }$.\ \ This follows
as\ \ ${\upsilon ( g) }^{-1}={\upsilon ( g) }^{\dagger }$ implies
$-i \upsilon ^{\prime }( X) ={(i \upsilon ^{\prime }( X) )}^{\dagger
}$ and hence $\upsilon ^{\prime }( X) ={\upsilon ^{\prime }( X)
}^{\dagger }$.}.\ \ Schur's lemma states that the representation
of the central generators are a multiple of the identity for irreducible
representations and so $\widehat{I}=\lambda  \text{\boldmath $1$}$ where
$\lambda \in \mathbb{R}\backslash \{0\}$.\ \ With $\lambda =1$,\ \ these
are the Heisenberg commutation relations given in (1).\ \ 
\subsection{Symmetry of Physical States}

A basic assumption of quantum mechanics is that the Heisenberg commutation
relations (1) are valid when acting on any physical state. Physically
observable probabilities are given by the square of the modulus
of the states. Therefore, physical states in quantum mechanics are
rays $\Psi $ that are equivalence classes of states $|\psi \rangle
$ in the Hilbert space that are equal up to a phase \cite{wigner},
\cite{Weinberg1}, 
\begin{equation}
\Psi =\left[ \left| \psi \right\rangle  \right] , \left| \widetilde{\psi
}\right\rangle  \simeq \left| \psi \right\rangle  \ \ \ \ \left|
\widetilde{\psi }\right\rangle  =e^{i \theta } \left| \psi \right\rangle
.
\end{equation}

\noindent The square of the modulus is the same for any representative
state in the ray, 
\begin{equation}
P( \alpha \rightarrow \beta ) =|\left( \Psi _{\beta },\Psi _{\alpha
}\right) |^{2}=|\left\langle  {\widetilde{\psi }}_{\beta }|{\widetilde{\psi
}}_{\alpha }\right\rangle  |^{2}=|\left\langle  \psi _{\beta }|\psi
_{\alpha }\right\rangle  |^{2}.
\end{equation}

Symmetry transformations between physical states (i.e. rays $\Psi$)
are given by operators $U$ that leave invariant the square of modulus,
\begin{equation}
|\left( U \Psi _{\beta },U \Psi _{\alpha }\right) |^{2}=|\left(
\Psi _{\beta },\Psi _{\alpha }\right) |^{2}.
\end{equation}

\noindent These transformations $U$ are the representation of a
group in the space $U( \mathrm{H}) $ of linear or anti-linear operators
on $\text{\boldmath $\mathrm{H}$}$ 
\begin{equation}
\varrho :\mathcal{G}\rightarrow U( \mathrm{H}) : g\mapsto U=\varrho
( g) .
\end{equation}

\noindent This\ \ operator also acts on any representative in the
equivalence class of states that defines the ray, 
\begin{equation}
\Psi ^{\prime }=U \Psi ,\ \ \left| \psi ^{\prime }\right\rangle
=U\left| \psi \right\rangle  .
\end{equation}

\noindent Theorem 2 in Appendix A states that any representation
of a Lie group \cite{bargmann}, \cite{mackey2} that leaves invariant
the square of the modulus is always equivalent to a linear unitary
or anti-linear, anti-unitary operator mapping the Hilbert space
$\text{\boldmath $\mathrm{H}$}$ into itself. Furthermore, if the
Lie group is connected\footnote{In this paper, a connected group
is abbreviation for a group for which every element is connected
by a continuous path to the identity element. }, it is always equivalent
to a linear unitary operator.\ \ 

The representations $\varrho $ are referred to as projective representations.\ \ If
$\mathcal{G}$ is a connected Lie group, the fundamental Theorem
3 states that these projective representations are equivalent to
the ordinary unitary representations $\upsilon $ of the central
extension $\widecheck{\mathcal{G}}$ of $\mathcal{G}$.

We seek the maximal group with projective representations that preserve
the Heisenberg commutation relations.\ \ As the Heisenberg commutation
relations are a faithful unitary representation of the Lie algebra
of the Weyl-Heisenberg group, the group we seek must be a subgroup
of the automorphism group of the Weyl-Heisenberg algebra.\ \ As
the Weyl-Heisenberg group is simply connected, the automorphism
group of the algebra is equivalent to the automorphism group ${\mathcal{A}ut}_{\mathcal{H}(
n) }$ of the Weyl-Heisenberg group itself. 

Under the action of elements $g\in {\mathcal{A}ut}_{\mathcal{H}(
n) }$, the elements of the algebra transform to a new basis
\begin{equation}
{P^{\prime }}_{i}=g P_{i} g^{-1}, {Q^{\prime }}_{i}=g Q_{i} g^{-1},
I^{\prime }=g I g^{-1}.
\end{equation}

\noindent such that the form of the Lie algebra is preserved,
\begin{equation}
\left[ {P^{\prime }}_{i},{Q^{\prime }}_{i}\right] =\delta _{i,j}I^{\prime
} .%
\label{MG: prined WH algebra}
\end{equation}

\noindent The element $I^{\prime }$ is central and as $I$ spans
the center of the algebra, we must have $I^{\prime }=d I$ with $d\in
\mathbb{R}\backslash \{0\}$. Furthermore, the elements of the automorphism
group that preserves the center of the algebra, 
\begin{equation}
 I^{\prime }=g I g^{-1}=I,
\end{equation}

\noindent defines a subgroup. 

The group inner automorphisms of a group\ \ is isomorphic to the
group itself.\ \ The full group of automorphisms always contains
the group of inner automorphisms as a normal subgroup. For the case
of the Weyl-Heisenberg group, this means that the Weyl-Heisenberg
group is a normal subgroup of its automorphism group, $\mathcal{H}(
n) \subset {\mathcal{A}ut}_{\mathcal{H}( n) }$.\ \ \ 

The projective representations of ${\mathcal{A}ut}_{\mathcal{H}(
n) }$ are equivalent to the unitary representations $\upsilon $
of its central extension ${\widecheck{\mathcal{A}ut}}_{\mathcal{H}(
n) }$\ \ acting on a Hilbert space ${\text{\boldmath $\mathrm{H}$}}^{\upsilon
}$.\ \ If we restrict $\upsilon $ to the normal subgroup $\mathcal{H}(
n) $ of inner automorphisms, these are the unitary representations
of the Weyl-Heisenberg group, $\upsilon |_{\mathcal{H}( n) }=\xi
$.\ \ Therefore, the Hilbert space ${\text{\boldmath $\mathrm{H}$}}^{\xi
}$ is an invariant subspace of ${\text{\boldmath $\mathrm{H}$}}^{\upsilon
}$.\ \ The generators of the Weyl-Heisenberg group transform under
the action of elements $U=\upsilon ( g) $, $g\in {\mathcal{A}ut}_{\mathcal{H}(
n) }$ as
\begin{equation}
\begin{array}{l}
 \begin{array}{l}
 {{\widehat{P}}^{\prime }}_{i}=\upsilon ^{\prime }( {P^{\prime }}_{i})
=\upsilon ^{\prime }( g P_{i}g^{-1}) =\upsilon ( g) \xi ^{\prime
}( P_{i}) {\upsilon ( g) }^{-1}=U {\widehat{P}}_{i}U^{-1}, \\
 {{\widehat{Q}}^{\prime }}_{i}=\upsilon ^{\prime }( {Q^{\prime }}_{i})
=\upsilon ^{\prime }( g Q_{i}g^{-1}) =\upsilon ( g) \xi ^{\prime
}( Q_{i}) {\upsilon ( g) }^{-1}=U {\widehat{Q}}_{i}U^{-1}, \\
 {\widehat{I}}^{\prime }=\upsilon ^{\prime }( I^{\prime }) =\upsilon
^{\prime }( g I g^{-1}) =\upsilon ( g) \xi ^{\prime }( I) {\upsilon
( g) }^{-1}=U \widehat{I}U^{-1},
\end{array}
\end{array}
\end{equation}

For the faithful representation $\upsilon $, the commutation relations
for the transformed generators are, using (15), 
\begin{equation}
\left[ {{\widehat{P}}^{\prime }}_{i},{{\widehat{Q}}^{\prime }}_{i}\right]
=i\ \ \delta _{i,j}{\widehat{I}}^{\prime }=i \lambda ^{\prime } \delta
_{i,j}\text{\boldmath $1$}
\end{equation}

\noindent where ${\widehat{I}}^{\prime }=d\widehat{I}$ and so $\lambda ^{\prime
}=d \lambda $.\ \ Now, as we have noted, the $\lambda $ label the
faithful irreducible representations of the Weyl-Heisenberg group.
Furthermore, the physical cases corresponds to the choice $\lambda
=1$.\ \ This must also be true for the transformed operators and
therefore $\lambda ^{\prime }=1$ and so ${\widehat{I}}^{\prime }=\widehat{I}$\ \ with
$d=1$.\ \ That is, the projective representation of the symmetry
group of the Heisenberg commutation relations leaves the representation
of the center of the Weyl-Heisenberg group invariant.\ \ As the
representation is faithful, the symmetry group also must leave the
central generator of the Weyl-Heisenberg algebra invariant, $I^{\prime
}=I$. 

Therefore, the maximal group of symmetries of the Heisenberg commutation
relations are the projective representation of the subgroup of the
automorphism group of the Weyl-Heisenberg group that leaves the
central generator $I$ invariant. 

The problem that this paper addresses is to determine the explicitly
this symmetry group and its projective representations. We will
show that the automorphism group of the Weyl-Heisenberg group is
\cite{folland}\ \ \ 
\begin{equation}
{\mathcal{A}ut}_{\mathcal{H}( n) }\simeq \mathcal{D}\otimes _{s}\mathcal{H}\overline{\mathcal{S}p}(
2n) ,%
\label{MG: Original Aut group def}
\end{equation}

\noindent where
\begin{equation}
\mathcal{H}\overline{\mathcal{S}p}( 2n) \simeq \overline{\mathcal{S}p}(
2n) \otimes _{s}\mathcal{H}( n) ,\ \ \ \ \ \mathcal{D}\simeq \left(
\mathbb{R}\backslash \left\{ 0\right\} ,\times \right) ,
\end{equation}

\noindent where $\mathcal{D}$ is the reals excluding $\{0\}$ viewed
as a group under multiplication, $\mathcal{D}\simeq (\mathbb{R}\backslash
\{0\},\times )$.\ \ We will show that the subgroup of the automorphism
group that leaves the central generator $I$ invariant is\ \ \ 
\begin{equation}
\mathcal{H}\overline{\mathcal{S}p}( 2n) .%
\label{MG: ce of symmmetry group}
\end{equation}

\noindent The group $\mathcal{H}\overline{\mathcal{S}p}( 2n) $ is
connected and is the central extension of the Inhomogeneous group,
$\mathcal{H}\overline{\mathcal{S}p}( 2n) \simeq \mathcal{I}\widecheck{\mathcal{S}p}(
2n) $ that is defined by the short exact sequence
\begin{equation}
e\rightarrow \mathbb{Z}\otimes \mathcal{A}( 1) \rightarrow \mathcal{H}\overline{\mathcal{S}p}(
2n) \rightarrow \mathcal{I}\mathcal{S}p( 2n) \rightarrow e.%
\label{MG: short exact sequence for hsp}
\end{equation}

\noindent $\mathbb{Z}$ is the center of $\overline{\mathcal{S}p}(
2n) $ and $\mathcal{A}( 1) $ is the center of $\mathcal{H}( n) $.\ \ $\mathcal{I}\mathcal{S}p(
2n) $ is the inhomogeneous symplectic group familiar from classical
Hamiltonian mechanics,
\begin{equation}
\mathcal{I}\mathcal{S}p( 2n) \equiv \mathcal{S}p( 2n) \otimes _{s}\mathcal{A}(
2n) .%
\label{MG: ISp}
\end{equation}

To establish the above results we start by reviewing the Weyl-Heisenberg
group. We then derive its automorphism group and the subgroup that
leaves the center of the Weyl-Heisenberg group invariant. This is
the maximal symmetry group. The projective representations of this
symmetry group are equivalent to the unitary representations of
its central extension.\ \ We use the Mackey theorems to compute
the unitary irreducible representations of the symmetry group from
first principles. (As the symmetry group contains the Weyl-Heisenberg
group as normal subgroup, this first requires the computation of
the faithful unitary irreducible representations of the Weyl-Heisenberg
group itself using the Mackey theorems.)\ \ We will enumerate and
comment on the degenerate cases.\ \ \ \ 
\section{The symmetry group}

In this section, we review basic properties of the Weyl-Heisenberg
group and determine its automorphism group.\ \ We then determine
the subgroup leaving the center of the Weyl-Heisenberg group invariant
and study certain of its properties. 
\subsection{The Weyl-Heisenberg group}

The Weyl-Heisenberg Lie group is defined to be the semi-direct product
of two abelian groups of the form given in (2). We first verify
that these group product (3) and inverse (4) relations result in
the semidirect product of this form.\ \ First, the group product
and inverse (3-4) enable us to identify the abelian subgroups
\begin{equation}
\Upsilon ( 0,q,\iota ) \in \mathcal{A}( n+1) , \Upsilon ( p,0,0)
\in \mathcal{A}( n) .%
\label{MG: WH normal q i}
\end{equation}

\noindent where again $p,q\in \mathbb{R}^{n}$ and $\iota \in \mathbb{R}$.\ \ These
subgroups satisfy the group product and inverse relations
\begin{gather}
\Upsilon ( 0,q^{\prime },\iota ^{\prime }) \Upsilon ( 0,q,\iota
) =\Upsilon ( 0,q^{\prime }+q,\iota +\iota ^{\prime }) ,\ \ {\Upsilon
( 0,q,\iota ) }^{-1}=\Upsilon ( 0,-q,-\iota ) 
\\\Upsilon ( p^{\prime },0,0) \Upsilon ( p,0,0) =\Upsilon ( p^{\prime
}+p,0,0) , {\Upsilon ( p,0,0) }^{-1}=\Upsilon ( -p,0,0) .%
\label{MG: Heisenberg group product}
\end{gather}

\noindent Additional abelian subgroups are likewise given by
\begin{equation}
\Upsilon ( p,0,\iota ) \in \mathcal{A}( n+1) ,\ \ \Upsilon ( 0,q,0)
\in \mathcal{A}( n+1) %
\label{MG: WH normal p i}
\end{equation}

We calculate the inner automorphisms of the group using (3-4) to
be\footnote{We always use $\varsigma $ to define the similarity
map $\varsigma _{g}h\equiv g h g^{-1}$ in what follows.}
\begin{equation}
\begin{array}{ll}
 \varsigma _{\Upsilon ( p^{\prime },q^{\prime },\iota ^{\prime })
}\Upsilon ( p,q,\iota )  & =\Upsilon ( p^{\prime },q^{\prime },\iota
^{\prime }) \Upsilon ( p,q,\iota ) {\Upsilon ( p^{\prime },q^{\prime
},\iota ^{\prime }) }^{-1} \\
  & =\Upsilon ( p,q,\iota +p^{\prime }q-q^{\prime }\cdot p) .
\end{array}%
\label{MG: WH inner aut unpolarized}
\end{equation}

\noindent In particular, note that for each of the choices of the
subgroups 
\begin{gather}
\varsigma _{\Upsilon ( p^{\prime },q^{\prime },\iota ^{\prime })
}\Upsilon ( 0,q,\iota ) =\Upsilon ( 0,q,\iota +p^{\prime }q) ,%
\label{MG: WH inner aut q t}
\\\varsigma _{\Upsilon ( p^{\prime },q^{\prime },\iota ^{\prime
}) }\Upsilon ( p,0,\iota ) =\Upsilon ( p,0,\iota -q^{\prime }\cdot
p) .%
\label{MG: WH inner aut p i}
\end{gather}

\noindent This means that both of the $\mathcal{A}( n+1) $ subgroups
given in (24), (27) are normal subgroups.\ \ Another special case
of (3) is 
\begin{equation}
\varsigma _{\Upsilon ( 0,0,\iota ^{\prime }) }\Upsilon ( p,q,\iota
) =\Upsilon ( p,q,\iota ) .%
\label{MG: center of H}
\end{equation}

\noindent and therefore the elements $\Upsilon ( 0,0,\iota ^{\prime
}) $ commute with all elements of the group. Furthermore, these
are the only elements that commute with all other elements of the
group.\ \ Therefore the $\mathcal{A}( 1) $ group that is defined
by the elements $\Upsilon ( 0,0,\iota ) $ is the center of the group,
$\mathcal{Z}\simeq \mathcal{A}( 1) $.

The final step to verify that the group relations defined by (3-4)
results in the Weyl-Heisenberg group having the structure of a semidirect
product given in (2).\ \ We have already established that there
are two choices for the $\mathcal{A}( n) $ subgroup and $\mathcal{A}(
n+1) $ normal subgroup.\ \ It is clear in both cases that 
\begin{equation}
\mathcal{A}( n) \cap \mathcal{A}( n+1) =\text{\boldmath $e$}\text{\boldmath
$,$}
\end{equation}

\noindent as the identity $\Upsilon ( 0,0,0) $ is the only element
in both groups for both cases.\ \ It remains to show that $\mathcal{A}(
n+1) \mathcal{A}( n) \simeq \mathcal{H}( n) $.\ \ Using the group
product (3),\ \ for each of the cases (24), (27),\ \ this is
\begin{gather}
\Upsilon ( 0,q,\iota ) \Upsilon ( p,0,0) =\Upsilon ( p,q,\iota -\frac{1}{2}q\cdot
p) ,
\\\Upsilon ( p,0,\iota ) \Upsilon ( 0,q,0) =\Upsilon ( p,q,\iota
+\frac{1}{2}p\cdot q) .
\end{gather}

The map
\begin{equation}
\varphi ^{\pm }:\mathcal{H}( n) \rightarrow \mathcal{H}( n) : \Upsilon
( p,q,\iota )  \mapsto \Upsilon ^{\pm }( p,q,\iota ^{\pm })  =\Upsilon
( p,q,\iota \mp \frac{1}{2}p\cdot q) %
\label{MG: WH Isomorphism}
\end{equation}

\noindent is a homomorphism that is onto and the kernel is trivial.
Therefore, the map $\varphi ^{\pm }$ is an isomorphism and the Weyl-Heisenberg
group has the semidirect product structure given in (2) for either
of the choices of abelian subgroup given by (24), (27). 

 The Weyl-Heisenberg\ \ Lie group is a matrix group and may be realized
by the $2n+2$ dimensional square matrices\ \ 
\begin{equation}
\Upsilon ( p,q,\iota ) =\left( \begin{array}{llll}
 1_{n} & 0 & 0 & p \\
 0 & 1_{n} & 1 & q \\
 q^{\mathrm{t}} & -p^{\mathrm{t}} & 1 & 2\iota  \\
 0 & 0 & 0 & 1
\end{array}\right) .
\end{equation}

\noindent $1_{m}$ denotes the unit matrix in $m$ dimensions and
the t superscript denotes the transpose.\ \ The group multiplication
and inverse (3-4) are realized by matrix multiplication and inverse.\ \ 

The Lie algebra of the Weyl-Heisenberg group may be computed from
this matrix realization.\ \ The coordinates are nonsingular at the
origin and therefore, choosing the unpolarized form, the generators
are given by
\begin{equation}
Q_{i}=\frac{\partial }{\partial p^{i}} \Upsilon ( p,q,\iota ) |_{e}
, P_{i}=\frac{\partial }{\partial q^{i}} \Upsilon ( p,q,\iota )
|_{e}, I=\frac{\partial }{\partial \iota } \Upsilon ( p,q,\iota
) |_{e}.
\end{equation}

\noindent A general element of the algebra is then
\begin{equation}
W=p^{i}Q_{i}+q^{i}P_{i}+\iota  I.
\end{equation}

\noindent The nonzero commutation relations are,\ \ as expected,
\begin{equation}
\left[ P_{i},Q_{i}\right] =\delta _{i,j}I
\end{equation}

\noindent where $I$ is a central generator. 

It is convenient to also introduce the notation that combines the
$p,q\in \mathbb{R}^{n}$ into a single $2n$ tuple $z=(p,q)$, $z\in
\mathbb{R}^{2n}$.\ \ Then the group product and inverse are 
\begin{equation}
\Upsilon ( z^{\prime },\iota ^{\prime }) \Upsilon ( z,\iota ) =\Upsilon
( z^{\prime }+z,\iota +\iota -\frac{1}{2} {z^{\prime }}^{\mathrm{t}}\zeta
z) ,\ \ {\Upsilon ( z,\iota ) }^{-1}=\Upsilon ( -z,-\iota ) %
\label{MG: z Heisenberg group product}
\end{equation}

\noindent and the unpolarized matrix realization is
\begin{equation}
\Upsilon ( z,\iota ) =\left( -\begin{array}{lll}
 1_{2n} & 0 & z \\
 z^{\mathrm{t}}\zeta  & 1 & 2\iota  \\
 0 & 0 & 1
\end{array}\right) ,\ \ \zeta =\left( \begin{array}{ll}
 0 & 1_{n} \\
 -1_{n} & 0
\end{array}\right) .%
\label{MG: H element z notation}
\end{equation}

\noindent The Lie algebra has general element 
\begin{equation}
W( z,\iota ) =z^{\alpha }Z_{\alpha }+\iota  I,%
\label{MG: H alg basis}
\end{equation}

\noindent $\alpha , \beta ,...=1,...2n\text{}$ where the matrix
form of the algebra is
\begin{equation}
W( z,\iota ) =\left( -\begin{array}{lll}
 0 & 0 & z \\
 z^{\mathrm{t}}\zeta  & 0 & 2\iota  \\
 0 & 0 & 0
\end{array}\right) , %
\label{MG: H algebra element z notation}
\end{equation}

\noindent The generators satisfy the nonzero commutation relations
\begin{equation}
\left[ Z_{\alpha },Z_{\beta }\right] =\zeta _{\alpha ,\beta }I.
\end{equation}
\subsection{The automorphism group of the Weyl-Heisenberg group}

The automorphism group of a group $\mathcal{G}$ is the maximal group
for which $\mathcal{G}$ is a normal subgroup. We have established
in the previous section that the Weyl-Heisenberg group is a simply
connected matrix group and this enables us to prove the following
theorem.\ \ \ \ \ 
\begin{theorem}

The automorphism group of the Weyl-Heisenberg group $\mathcal{H}(
n) $ is\label{PH: theorem: WH Automorphism theorem}
\end{theorem}
\begin{equation}
{\mathcal{A}ut}_{\mathcal{H}( n) }\simeq \mathcal{D}\otimes _{s}\overline{\mathcal{S}p}(
2n) \otimes _{s}\mathcal{H}( n) .%
\label{MG: aut D semi HSp}
\end{equation}

\noindent $\mathcal{H}( n) $ is the Weyl-Heisenberg group, $\overline{\mathcal{S}p}(
2n) $ is the cover of the real symplectic group that leaves invariant
a real skew symmetric form and $\mathcal{D}$ is the reals excluding
zero viewed as a group under multiplication $\mathcal{D}\simeq (\mathbb{R}\backslash
\{0\},\times )$. 

As the Weyl-Heisenberg group is simply connected, Theorem 7\ \ states
that\ \ the automorphism group of its algebra and group are equivalent.\ \ We
can therefore establish the result by determining the maximal group
for which its elements\ \ $\Omega $ satisfy
\begin{equation}
\varsigma _{\Omega }W=\Omega  W \Omega ^{-1}=W^{\prime }.%
\label{MG: aut of alg}
\end{equation}

\noindent $W,W^{\prime }$ are general elements of the algebra of
the Weyl-Heisenberg group (42). The most general transformation
between a primed and unprimed basis is
\begin{equation}
{Z^{\prime }}_{\alpha }= a_{\alpha }^{\beta }Z_{\beta }+x_{\alpha
} I,\ \ I^{\prime }=c^{\alpha }Z_{\alpha }+d I.%
\label{MG: General aut on alg gen}
\end{equation}

\noindent The commutator [${Z^{\prime }}_{\alpha },I^{\prime }]=0$
requires $c^{\alpha }=0$ so that $I^{\prime }=d I$ with $d\in \mathbb{R}\backslash
\{0\}$.\ \ Next, 
\begin{equation}
\begin{array}{rl}
 \zeta _{\alpha ,\beta }I^{\prime }=\left[ {Z^{\prime }}_{\alpha
},{Z^{\prime }}_{\beta }\right]  & =\left[ a_{\alpha }^{\kappa }Z_{\kappa
}+x_{\alpha },a_{\beta }^{\kappa }Z_{\kappa }+x_{\beta }\right]
\\
  & =\frac{1}{d}a_{\alpha }^{\delta }a_{\beta }^{\gamma }\zeta _{\delta
,\gamma }I^{\prime }.
\end{array}
\end{equation}

\noindent This has the solution $a_{\alpha }^{\beta }=\delta  \Sigma
_{\alpha }^{\beta }$ and\ \ $d=\delta ^{2}$.\ \ Therefore, for $W(
z,\iota ) =z^{\kappa }Z_{\kappa }+\iota  I$ we have
\begin{equation}
W( z^{\prime },\iota ^{\prime }) =\varsigma _{\Omega }W( z,\iota
) ={z^{\prime }}^{\kappa }Z_{\kappa }+\iota ^{\prime } I
\end{equation}

\noindent with\cite{folland}
\begin{equation}
z^{\prime }=\delta  \Sigma  z, \iota ^{\prime }=\iota  \delta ^{2}+z\cdot
x.%
\label{MG: general form of W prime}
\end{equation}

To determine the group with elements $\Omega $ that satisfies (46,
50), we can use the matrix realization of the algebra given in (43).\ \ As
$\Omega $ is nonsingular, (46) is equivalent to\ \ 
\begin{equation}
\Omega  W( z,\iota )  =W( z^{\prime },\iota ^{\prime }) \Omega .%
\label{MG: aut of alg W}
\end{equation}

\noindent where $\Omega $ is a $2n+2$ dimensional square matrix.
We can write $\Omega $ in terms of the submatrices
\begin{equation}
 \Omega =\left( \begin{array}{lll}
 a & c & z \\
 f & d & j \\
 g & h & \epsilon 
\end{array}\right) 
\end{equation}

\noindent where $j,d,r,h,e\in \mathbb{R}$, $c,w,f,g\in \mathbb{R}^{2n}$
and $a$ is a $2n$ dimensional square submatrices and then solve
(51) to obtain 
\begin{equation}
 \Omega ( \delta ,\Sigma ,z,\iota ) =\left( \begin{array}{lll}
  \delta  \Sigma  & 0 & z \\
 - \delta  z^{\mathrm{t}}\zeta  \Sigma  &  \delta ^{2} & 2 \iota
\\
 0 & 0 & 1
\end{array}\right) 
\end{equation}

\noindent where $z\in \mathbb{R}^{2n}$, $\delta \in \mathcal{D}\equiv
\mathbb{R}\backslash \{0\}$,\ \ $\iota \in \mathbb{R}$ and $\Sigma
^{\mathrm{t}}\zeta  \Sigma =\zeta $ and so $\Sigma \in \mathcal{S}p(
2n) $.\ \ 

Direct matrix multiplication shows that the elements $\Omega ( \delta
,\Sigma ,w,r) $ define a group that we call ${\mathrm{aut}}_{\mathcal{H}(
n) }$ with product and inverse
\begin{gather}
\begin{array}{rl}
 \Omega ( \delta ^{{\prime\prime}},\Sigma ^{{\prime\prime}},z^{{\prime\prime}},\iota
^{{\prime\prime}})  & =\Omega ( \delta ^{\prime },\Sigma ^{\prime
},z^{\prime },\iota ^{\prime }) \Omega ( \delta ,\Sigma ,z,\iota
)  \\
  & =\Omega ( \delta ^{\prime }\delta ,\Sigma ^{\prime }\Sigma ,z^{\prime
}+\delta ^{\prime }\Sigma ^{\prime }z, \iota ^{\prime }+ {\delta
^{\prime }}^{2}\iota -\frac{1}{2}\delta ^{\prime } z^{\prime t}\zeta
\Sigma ^{\prime }z) ,
\end{array}%
\label{MG: aut group product}
\\{\Omega ( \delta ,\Sigma ,z,\iota ) }^{-1}=\Omega ( \delta ^{-1},\Sigma
^{-1},- \delta ^{-1}\Sigma ^{-1}z,-\delta ^{-2}\iota ) ,%
\label{MG: aut group inverse}
\end{gather}

\noindent where the identity element is $e=\{1,1_{2n},0,0\}$.\ \ From
these relations, we can explicitly compute that automorphisms of
the algebra given in (46)
\begin{equation}
W( z^{\prime },\iota ^{\prime }) =\varsigma _{\Omega ( \delta ,\Sigma
,z^{{\prime\prime}},\iota ^{{\prime\prime}}) }W( z,\iota ) 
\end{equation}

\noindent where
\begin{equation}
z^{\prime }=\delta  \Sigma  z, \iota ^{\prime }=\iota  \delta ^{2}-\delta
{z^{{\prime\prime}}}^{\mathrm{t}}\zeta  \Sigma \cdot z.%
\label{MG: general aut of WH algebra}
\end{equation}

\noindent Comparing with the general expression given in (50), they
are equivalent where we identify\ \ $x=\delta  {z^{{\prime\prime}}}^{\mathrm{t}}\zeta
\Sigma $.\ \ As $\det  \Sigma \neq 0 \mathrm{and} \delta \neq 0$,
there is a bijection between values of $x$ and $z^{{\prime\prime}}$.

Using these relations, the next step is to show that the group ${\mathrm{aut}}_{\mathcal{H}(
n) }$ has the form of a semidirect product\footnote{The definition
of a semidirect product is reviewed in Definition 1 in Appendix
A.}
\begin{equation}
{\mathrm{aut}}_{\mathcal{H}( n) }\simeq \left( \mathcal{D}\otimes
\mathcal{S}p( 2n) \right) \otimes _{s}\mathcal{H}( n) .%
\label{MG: aut D direct Sp semi H}
\end{equation}

First, using the group product and inverse (54-55), we can establish
that $\mathcal{D}$, $\mathcal{S}p( 2n) $ and $\mathcal{H}( n) $
are subgroups of ${\mathrm{aut}}_{\mathcal{H}( n) }$ with elements
\begin{equation}
\ \ \begin{array}{l}
 \Omega ( \delta ,1_{2n},0,0) \in \mathcal{D}, \\
 \Omega ( 1,\Sigma ,0,0) \simeq \Sigma \in \mathcal{S}p( 2n)  \\
 \Omega ( 1,1_{2n},z,\iota ) =\Upsilon ( z,\iota ) \in \mathcal{H}(
n) 
\end{array}
\end{equation}

The direct product $\mathcal{D}\otimes \mathcal{S}p( 2n) $ is immediately
established from the special case of the group multiplication (54)
\begin{equation}
\begin{array}{rl}
 \Omega ( \delta ,\Sigma ,0,0)  & =\Omega ( \delta ,1_{2n},0,0)
\Omega ( 1,\Sigma ,0,0)  \\
  & =\Omega ( 1,\Sigma ,0,0) \Omega ( \delta ,1_{2n},0,0) ,
\end{array}%
\label{MG: aut group product}
\end{equation}

\noindent The semidirect product in (58) is established by first
noting that\ \ 
\begin{equation}
\left( \mathcal{D}\otimes \mathcal{S}p( 2n) \right) \cap \mathcal{H}(
n) \simeq \left\{ \Omega ( \delta ,\Sigma ,0,0) \right\} \cap \left\{
\Omega ( 1,1_{2n},z,\iota ) \right\} \simeq e,
\end{equation}

\noindent  Then, using the group product (54), 
\begin{equation}
\Omega ( 1,1_{2n},z,\iota ) \Omega ( \delta ,\Sigma ,0,0) =\Omega
( \delta ,\Sigma ,z,\iota ) .
\end{equation}

\noindent Direct computation using (54-55) shows that the Weyl-Heisenberg
subgroup $\mathcal{H}( n) $ is a normal subgroup with the automorphisms
given by
\begin{equation}
\varsigma _{\Omega ( \delta ^{\prime },\Sigma ^{\prime },z^{\prime
},\iota ^{\prime }) }\Upsilon ( z,\iota ) =\Upsilon ( \delta ^{\prime
} \Sigma ^{\prime }z,{\delta ^{\prime }}^{2} \iota -\delta ^{\prime
}{z^{\prime }}^{\mathrm{t}}\zeta  \Sigma ^{\prime }z) .%
\label{MG: automorphisms of aut}
\end{equation}

\noindent This establishes that ${\mathrm{aut}}_{\mathcal{H}( n)
}$ has the semidirect product form given in (58).\ \ The right associative
property of the semidirect product allows\ \ this to be written
as 
\begin{equation}
\begin{array}{rl}
 {\mathrm{aut}}_{\mathcal{H}( n) } & \simeq \left( \mathcal{D}\otimes
\mathcal{S}p( 2n) \right) \otimes _{s}\mathcal{H}( n)  \\
  & \simeq \mathcal{D}\otimes _{s}\mathcal{H}\mathcal{S}p( 2n) 
\end{array}%
\label{MG: aut D semi HSp}
\end{equation}

\noindent where $\mathcal{H}\mathcal{S}p( 2n) $ is a semidirect
product of the form
\begin{equation}
\mathcal{H}\mathcal{S}p( 2n) \simeq \mathcal{S}p( 2n) \otimes _{s}\mathcal{H}(
n) %
\label{MG: HSp is Sp semi H}
\end{equation}

This the local characterization of the automorphism group.\ \ It
remains to consider any global topological properties that could
result in a larger group that behaves the same locally.\ \ 

The group $\mathcal{D}$ may be written as the direct product $\mathcal{D}\simeq
\mathbb{Z}_{2}\otimes \mathcal{D}^{+}$ where $\mathcal{D}^{+}\simeq
(\mathbb{R}^{+},\times )$ is the positive reals considered as a
group under multiplication.\ \ $\mathbb{Z}_{2}$ is the discrete
group with two elements $\{\pm 1\}$. $\mathcal{D}^{+}$ is simply
connected but $\mathcal{D}$ has two components,\ \ $\mathcal{D}/\mathcal{D}^{+}\simeq
\mathbb{Z}_{2}$. Therefore, the connected component of the group
is 
\begin{equation}
{\mathrm{aut}}_{\mathcal{H}( n) }^{c}\simeq \mathcal{D}^{+}\otimes
_{s}\mathcal{S}p( 2n) \otimes _{s}\mathcal{H}( n) .%
\label{MG: connected aut D semi HSp}
\end{equation}

\noindent $\mathcal{H}( n) $ and $\mathcal{D}^{+}$ are simply connected
and $\mathcal{S}p( 2n) $ is connected with fundamental group $\mathbb{Z}$.\ \ Its
simply connected universal cover is denoted $\overline{\mathcal{S}p}(
2n) $ with
\begin{equation}
\pi :\overline{\mathcal{S}p}( 2n) \rightarrow \mathcal{S}p( 2n)
:\overline{\Sigma }\rightarrow \Sigma =\pi ( \overline{\Sigma })
, \ker  \pi  \simeq \mathbb{Z}.
\end{equation}

\noindent Therefore, by the universal covering theorem,
\begin{equation}
{\mathcal{A}ut}_{\mathcal{H}( n) }^{c}\simeq {\overline{\mathrm{aut}}}_{\mathcal{H}(
n) }^{c}\simeq \mathcal{D}^{+}\otimes _{s}\overline{\mathcal{S}p}(
2n) \otimes _{s}\mathcal{H}( n) ,%
\label{MG: connected aut D semi HSp}
\end{equation}

\noindent is well defined and unique with the following group product
and inverse
\begin{gather}
\begin{array}{rl}
 \Omega ( \delta ^{{\prime\prime}},\Sigma ^{{\prime\prime}},z^{{\prime\prime}},\iota
^{{\prime\prime}})  & =\Omega ( \delta ^{\prime },{\overline{\Sigma
}}^{\prime },z^{\prime },\iota ^{\prime }) \Omega ( \delta ,\overline{\Sigma
},z,\iota )  \\
  & =\Omega ( \delta ^{\prime }\delta ,{\overline{\Sigma }}^{\prime
}\overline{\Sigma },z^{\prime }+\delta ^{\prime }\Sigma ^{\prime
}z, \iota ^{\prime }+ {\delta ^{\prime }}^{2}\iota -\frac{1}{2}\delta
^{\prime } z^{\prime t}\zeta  \Sigma ^{\prime }z) ,
\end{array}%
\label{MG: cover group product}
\\{\Omega ( \delta ,\Sigma ,z,\iota ) }^{-1}=\Omega ( \delta ^{-1},{\overline{\Sigma
}}^{-1},- \delta ^{-1}\Sigma ^{-1}z,-\delta ^{-2}\iota ) .%
\label{MG: cover group inverse}
\end{gather}

\noindent where $z\in \mathbb{R}^{2n}$, $\delta \in \mathcal{D}^{+}$,\ \ $\iota
\in \mathbb{R}$ and\ \ $\overline{\Sigma }\in \overline{\mathcal{S}p}(
2n) $.\ \ Note that in these expressions $\Sigma =\pi ( \overline{\Sigma
}) $$\text{}$. The expression for automorphisms of the Weyl-Heisenberg
subgroup remains the same as given in (63).

The cover of a disconnected group may be defined to be the central
extension of the group with a discrete central group. The problem
is, that in general, this does not give a unique cover and so this
must be checked on a case by case basis. This is discussed in Appendix
C where we show that
\begin{equation}
{\mathcal{A}ut}_{\mathcal{H}( n) }\simeq {\mathrm{aut}}_{\mathcal{H}(
n) }\simeq \mathcal{D}\otimes _{s}\overline{\mathcal{S}p}( 2n) \otimes
_{s}\mathcal{H}( n) %
\label{MG: connected aut D semi HSp}
\end{equation}

\noindent is unique and well defined. It has the group product and
inverse given in (69-70) where now $\delta \in \mathcal{D}$.\ \ Again,
the automorphisms of the Weyl-Heisenberg subgroup remains the same
as given in (63). 

The group ${\mathcal{A}ut}_{\mathcal{H}( n) }$ is the largest group
that the topological properties admit that is homomorphic to ${\mathrm{aut}}_{\mathcal{H}(
n) }$ and therefore we completed the proof of Theorem 1.\ \ 
\subsection{Subgroup of automorphism group with invariant center}

The action of the automorphism group on the algebra is given in\ \ (57).\ \ Invariance
of the central element requires $\delta =1$ which is the unit element
for $\mathcal{D}^{+}$.\ \ Thus the maximal symmetry group that leaves
the center of the Weyl-Heisenberg algebra invariant is $\mathcal{H}\overline{\mathcal{S}p}(
2n) $. 

As given in (22), the central extension of $\mathcal{H}\mathcal{S}p(
2n) $ is equivalent to the central extension of the inhomogeneous
symplectic group familiar from classical mechanics.\ \ \ 
\begin{equation}
\mathcal{H}\widecheck{\mathcal{S}p}( 2n) \simeq \mathcal{H}\overline{\mathcal{S}p}(
2n) \simeq \mathcal{I}\widecheck{\mathcal{S}p}( 2n) 
\end{equation}

This is a very remarkable fact.\ \ The central extension of the
$\mathcal{A}( 2n) $ is generally $n( 2n-1) $ dimensional. However,
because it is a subgroup of $\mathcal{I}\widecheck{\mathcal{S}p}( 2n)
$, the Lie algebra relations with the symplectic group constrain
the central extension of the abelian normal subgroup to be precisely
the one dimensional extension that is the Weyl-Heisenberg group.\ \ 

The group product and inverse are given by (69-70) with $\delta
=1$.
\subsubsection{Symplectic group factorization }

The defining condition for the real symplectic group $\mathcal{S}p(
2n) $ is 
\begin{equation}
\Sigma ^{\mathrm{t}}\zeta  \Sigma =\zeta %
\label{MG: symplectic condition}
\end{equation}

\noindent where $\zeta $ is the symplectic matrix defined in (41).
Matrix realizations of elements of the real symplectic group may
be written as 
\begin{equation}
\Sigma =\left( \begin{array}{ll}
 \Sigma _{1} & \Sigma _{2} \\
 \Sigma _{3} & \Sigma _{4}
\end{array}\right) 
\end{equation}

\noindent where $\Sigma _{a}$ , $a=1,..,4$ are $n\times n$ submatrices.\ \ 
The symplectic condition (73) immediately results in the relations
\begin{equation}
\begin{array}{l}
 \Sigma _{1}^{\mathrm{t}}\Sigma _{4}-\Sigma _{3}^{\mathrm{t}}\Sigma
_{2}=1_{n}, \\
 \Sigma _{1}^{\mathrm{t}} \Sigma _{3}={\left( \Sigma _{1}^{\mathrm{t}}
\Sigma _{3}\right) }^{\mathrm{t}}, \\
 \Sigma _{2}^{\mathrm{t}}\Sigma _{4}={\left( \Sigma _{2}^{\mathrm{t}}
\Sigma _{4}\right) }^{\mathrm{t}}.
\end{array}%
\label{MG: symplectic block matrix conditions}
\end{equation}

 A matrix realization of a Lie group is a coordinate system. As
$\operatorname{Det}( \Sigma ) =1$, it follows that the determinate
of at least one of the $\Sigma _{a}$, $a=1,...,4$,\ \ must be nonzero.\ \ These
correspond to different coordinate patches for the manifold underlying
the symplectic group. Assume $\operatorname{Det}( \Sigma _{1}) \neq
0$.\ \ Then \cite{degosson}, 
\begin{equation}
\Sigma ( \alpha ,\beta ,\gamma ) =\left( \begin{array}{ll}
 1_{n} & 0 \\
 \gamma  & 1_{n}
\end{array}\right) \left( \begin{array}{ll}
 \alpha ^{-1} & 0 \\
 0 & \alpha ^{\mathrm{t}}
\end{array}\right) \left( \begin{array}{ll}
 1_{n} & \beta  \\
 0 & 1_{n}
\end{array}\right) ,%
\label{MG: Sigma matrix factors f}
\end{equation}

\noindent where we define
\begin{equation}
\alpha ={\left( \Sigma _{1}\right) }^{-1}, \beta ={\left( \Sigma
_{1}\right) }^{-1}\Sigma _{2}, \gamma ={\Sigma _{3}( \Sigma _{1})
}^{-1}.
\end{equation}

\noindent It follows from (75) that $\beta =\beta ^{\mathrm{t}}$
and $\gamma =\gamma ^{\mathrm{t}}$. The matrix realizations of elements
of the symplectic group factor as
\begin{equation}
\Sigma ( \alpha ,\beta ,\gamma ) =\Sigma ^{-}( \gamma ) \Sigma \mbox{}^{\circ}(
\alpha ) \Sigma ^{+}( \beta ) %
\label{MG: Sigma matrix factors}
\end{equation}

\noindent where
\begin{equation}
\begin{array}{l}
 \Sigma \mbox{}^{\circ}( \alpha ) \equiv \Sigma ( \alpha ,1_{n},1_{n})
\in \mathcal{U}( n) , \\
 \Sigma ^{+}( \beta ) \equiv \Sigma ( 1_{n},\beta ,1_{n}) \in \mathcal{A}(
m) , \\
 \Sigma ^{-}( \gamma ) \equiv \Sigma ( 1_{n},1_{n},\gamma ) \in
\mathcal{A}( m) .
\end{array}
\end{equation}

\noindent and $m=\frac{n( n-1) }{2}$.\ \ Furthermore, note that
\begin{equation}
\zeta  \Sigma ^{-}( \gamma )  \zeta ^{-1}= \Sigma ^{+}( -\gamma
) . %
\label{MG: sigma gamma zeta tx}
\end{equation}

A similar argument applies if we instead assume $\operatorname{Det}(
\Sigma _{4}) \neq 0$. Both of these coordinate patches contain the
identity, $1_{2n}$ but neither contains the element $\zeta $.\ \ These
require us to consider the case with either $\Sigma _{2}$ or $\Sigma
_{3}$ to be assumed to be nonsingular.\ \ In this case, define
\begin{equation}
\widetilde{\Sigma }= \Sigma  \zeta ^{-1}=\left( \begin{array}{ll}
 \Sigma _{2} & -\Sigma _{1} \\
 \Sigma _{4} & -\Sigma _{3}
\end{array}\right) .
\end{equation}

\noindent The $\widetilde{\Sigma }$ also satisfy the symplectic condition
as\ \ 
\begin{equation}
\zeta =\Sigma ^{\mathrm{t}}\zeta  \Sigma =\zeta ^{\mathrm{t}}{\widetilde{\Sigma
}}^{\mathrm{t}}\zeta  \widetilde{\Sigma } \zeta  \Rightarrow {\widetilde{\Sigma
}}^{\mathrm{t}}\zeta  \widetilde{\Sigma }=\zeta .%
\label{MG: zeta factorization}
\end{equation}

\noindent This symplectic condition results in the identities
\begin{equation}
\begin{array}{l}
 \Sigma _{4}^{\mathrm{t}}\Sigma _{1}-\Sigma _{2}^{\mathrm{t}}\Sigma
_{3}=1_{n}, \\
 \Sigma _{2}^{\mathrm{t}} {\widetilde{\Sigma }}_{4}={\left( \Sigma _{2}^{\mathrm{t}}
\Sigma _{4}\right) }^{\mathrm{t}}, \\
 \Sigma _{1}^{\mathrm{t}}\Sigma _{3}={\left( \Sigma _{1}^{\mathrm{t}}
\Sigma _{3}\right) }^{\mathrm{t}}.
\end{array}
\end{equation}

We can now assume $\operatorname{Det}( \Sigma _{2}) \neq 0$ and
the analysis proceeds as before with 
\begin{equation}
\alpha ={\left( \Sigma _{2}\right) }^{-1}, \beta =-{\left( \Sigma
_{2}\right) }^{-1} \Sigma _{1}, \gamma =\Sigma _{4} {\left( \Sigma
_{2}\right) }^{-1},
\end{equation}

\noindent In this case the factorization must include the symplectic
matrix from (82)
\begin{equation}
\Sigma ( \alpha ,\beta ,\gamma ) =\Sigma ^{-}( \gamma ) \Sigma \mbox{}^{\circ}(
\alpha ) \Sigma ^{+}( \beta ) \zeta .%
\label{MG: Z symplectic factorization}
\end{equation}

\noindent Finally a similar argument applies for the coordinate
patch\ \ $\operatorname{Det}( \Sigma _{3}) \neq 0$. Both of these
coordinate patches contain the element $\zeta $ but do not contain
the identity 

The expressions (78) and (85)\ \ can be combined into a single expression
\begin{equation}
\Sigma ^{\epsilon }( \alpha ,\beta ,\gamma ) =\Sigma ^{-}( \gamma
) \Sigma \mbox{}^{\circ}( \alpha ) \Sigma ^{+}( \beta ) \zeta ^{
\epsilon }.%
\label{MG: symplectic factorization}
\end{equation}

\noindent where $\epsilon \in \{0,1\}$.
\subsubsection{Lie Algebra}

The Lie algebra of the symmetry group $\mathcal{H}\overline{\mathcal{S}p}(
2n) $ is the same as the Lie algebra of $\mathcal{H}\mathcal{S}p(
2n) $.\ \ It may be directly computed from its matrix realization.
It is convenient to use a basis for the algebra of the symplectic
group corresponding to the factorized form (78). Let the $A_{i,j}$
be the generators of the unitary subgroup with elements $\Sigma
( \alpha ) \in \mathcal{U}( n) $, and $B_{i,j}$ the generators of
the abelian subgroup with elements\ \ $\Sigma ( \beta ) \in \mathcal{A}(
m) $ and $C_{i,j}$ the generators of the abelian subgroup with elements
$\Sigma ( \gamma ) \in \mathcal{A}( m) $.\ \ \ The abelian generators
are symmetric, $B_{i,j}=B_{j,i}$ and $C_{i,j}=C_{j,i}$.\ \ A general
element is written as 
\begin{equation}
Z=\alpha ^{i,j}A_{i,j}+\beta ^{i,j}B_{i,j}+\gamma ^{i,j}C_{i,j}+p^{i}Q_{i}+q^{i}P_{i}+\iota
I.
\end{equation}

Straightforward computation shows that these generators of $\mathcal{S}p(
2n) $ satisfy the Lie algebra
\begin{equation}
\begin{array}{l}
 \left[ A_{i,j},A_{k,l}\right]  =\delta _{i,l}A_{j,k}-\delta _{j,k}A_{i,l},
\\
 \left[ A_{i,j},B_{k,l}\right]  =\delta _{j,k}B_{i,l}+\delta _{j,l}B_{i,k},
\\
 \left[ A_{i,j},C_{k,l}\right]  =-\delta _{i,k}C_{j,l}-\delta _{i,l}C_{k,j},
\\
 \left[ B_{i,j},C_{k}\right]  =\delta _{i,k}A_{j,l} +\delta _{i,l}A_{j,k}
+\delta _{j,k}A_{i,l} +\delta _{j,l}A_{i,k}.
\end{array}
\end{equation}

The nonzero commutators of the algebra of $\mathcal{H}\mathcal{S}p(
2n) $ are the above relations for the symplectic generators together
with the Weyl-Heisenberg generators are
\begin{equation}
\begin{array}{ll}
 \left[ A_{i,j},Q_{k}\right] = \delta _{j,k}Q_{i}, & \left[ C_{i,j},Q_{k}\right]
=\delta _{j,k}P_{i}+ \delta _{i,k}P_{j}, \\
 \left[ A_{i,j},P_{k}\right] = -\delta _{i,k}P_{j}, & \left[ B_{i,j},P_{k}\right]
= \delta _{j,k}Q_{i}+\delta _{i,k}Q_{j}, \\
 \left[ P_{i},Q_{j}\right] =\delta _{i,j}I. &  
\end{array}
\end{equation}

The symplectic generators may be realized in the enveloping algebra
up to a central element \cite{Low13}.\ \ This will be important
when we discuss the representations in Section 3.2. 
\begin{equation}
{\widetilde{A}}_{i,j}= Q_{i}P_{j},\ \ {\widetilde{B}}_{i,j}=Q_{i}Q_{j},
{\widetilde{C}}_{i,j}= P_{i}P_{j}.
\end{equation}

Clearly $B_{i,j}=B_{j,i}$ and $C_{i,j}=C_{j,i}$.\ \ Then, using
the Weyl-Heisenberg commutation relations (5), this defines the
commutation relations, up to the central element, $I$,
\begin{equation}
\begin{array}{l}
 \left[ {\widetilde{A}}_{i,j},{\widetilde{A}}_{k,l}\right]  =I( \delta _{i,l}{\widetilde{A}}_{j,k}-\delta
_{j,k}{\widetilde{A}}_{i,l}) , \\
 \left[ A_{i,j},{\widetilde{B}}_{k,l}\right]  =I( \delta _{j,k}{\widetilde{B}}_{i,l}+\delta
_{j,l}{\widetilde{B}}_{i,k}) , \\
 \left[ A_{i,j},{\widetilde{C}}_{k,l}\right]  =-I( \delta _{i,k}{\widetilde{C}}_{j,l}+\delta
_{i,l}{\widetilde{C}}_{k,j}) , \\
 \left[ {\widetilde{B}}_{i,j},{\widetilde{C}}_{k}\right]  =I( \delta _{i,k}{\widetilde{A}}_{j,l}
+\delta _{i,l}{\widetilde{A}}_{j,k} +\delta _{j,k}{\widetilde{A}}_{i,l}
+\delta _{j,l}{\widetilde{A}}_{i,k} ) .
\end{array}%
\label{MG: symplectic in enveloping algebra of H}
\end{equation}
\section{Quantum symmetry: Projective representations}

The projective representations of the maximal symmetry group $\mathcal{I}\mathcal{S}p(
2n) $ are equivalent to the ordinary unitary representations of
its central extension $\mathcal{H}\overline{\mathcal{S}p}( 2n) $.
These unitary irreducible representations may be determined using
the Mackey theorems for semidirect product groups. 

The first step in applying the Mackey theorem for semidirect products
is to determine the unitary irreducible representations of the Weyl-Heisenberg
normal subgroup. While these are well known, the method of constructing
them using the Mackey theorems applied to the semidirect product
of two abelian groups (2) does not appear to be as well known \cite{Major}.
We review this briefly in order to introduce the Mackey theorems
and also to show how the unitary irreducible representations of
the symmetry group can be constructed completely from first principles.
\subsection{Unitary irreducible representations of the Weyl-Heisenberg
group}\label{MG: Section: WH UIR}

The Mackey theorem for semidirect products with an abelian normal
subgroup are given in Theorem 10 in\ \ Appendix A \cite{mackey}.
We choose the normal subgroup (27) with elements $\Upsilon ( p,0,\iota
) \in \mathcal{A}( n+1) $. The unitary irreducible representations
$\xi $ of the abelian normal subgroup are the phases acting on the
Hilbert space ${\text{\boldmath $\mathrm{H}$}}^{\xi }=\mathbb{C}$
\begin{equation}
\left. \xi ( \Upsilon ( p,0,\iota ) ) |\phi \right\rangle  =e^{i(
\iota  \widehat{I} +p^{i}{\widehat{Q}}_{i}) }\overset{ }{\left. |\phi \right\rangle
=e^{i( \iota  \lambda  +p\cdot \alpha ) }\overset{ }{\left. |\phi
\right\rangle  }},\ \ \overset{ }{\left. |\phi \right\rangle  }\in
\mathbb{C}.%
\label{MG: abelian uir}
\end{equation}

\noindent The hermitian representation of the algebra has the eigenvalues
that are given by
\begin{equation}
 {\widehat{Q}}_{i}\overset{ }{\left. |\phi \right\rangle  }=\xi ^{\prime
}( Q_{i}) \left| \phi \right\rangle  =\alpha _{i}\overset{ }{\left.
|\phi \right\rangle  },\ \ \ \widehat{I}\overset{ }{\left. |\phi \right\rangle
}=\xi ^{\prime }( I) \left| \phi \right\rangle  =\lambda \overset{
}{\left. |\phi \right\rangle  },\ \ 
\end{equation}

\noindent where $\alpha \in \mathbb{R}^{n}$ and $\lambda \in \mathbb{R}$.\ \ The
characters\ \ $\xi _{\alpha ,\lambda }$ are parameterized by the
eigenvalues $\alpha ,\lambda $ and the equivalence classes that
are elements of the unitary dual, $[\xi _{\alpha ,\lambda }]\in
{\text{\boldmath $U$}}_{\mathcal{A}( n+1) }\simeq \mathbb{R}^{n+1}$.\ \ Each
equivalence class has the single element $[\xi _{\alpha ,\lambda
}]=\xi _{\alpha ,\lambda }$. 

The action of the elements $\Upsilon ( 0,q,0) \in \mathcal{A}( n)
$ of the\ \ homogeneous group on these representations is given
by the dual automorphisms 
\begin{equation}
\begin{array}{l}
 \begin{array}{l}
 \left. \left. \left( {\widehat{\varsigma }}_{\Upsilon ( 0,q,0) }\xi
_{\alpha ,\lambda }\right) \left( \Upsilon ( p,0,\iota ) \right)
|\phi \right\rangle  =\xi _{\alpha ,\lambda }( \varsigma _{\Upsilon
( 0,q,0) }\Upsilon ( p,0,\iota ) ) |\phi \right\rangle   \\
 =\xi _{\alpha -\lambda  q,\lambda }( \Upsilon ( p,0,\iota ) ) \overset{
}{\left. |\phi \right\rangle  }.
\end{array}
\end{array}
\end{equation}

\noindent In simplifying this expression, we have used (30) and
(92). The little group is the set of $\Upsilon ( 0,q,0) \in \mathcal{K}\mbox{}^{\circ}$
that satisfy the fixed point equation (134), 
\begin{equation}
{\widehat{\varsigma }}_{\Upsilon ( 0,q,0) }\xi _{ \alpha ,\lambda }=\xi
_{\alpha -\lambda  q,\lambda } =\xi _{\alpha ,\lambda }.
\end{equation}

\noindent The solution of the fixed point condition requires that
$\alpha -\lambda  q\equiv \alpha $. The $\lambda =0$ solution for
which the little group is $\mathcal{A}( n) $ is the degenerate case
corresponding to the homomorphism $\mathcal{H}( n) \rightarrow \mathcal{A}(
2n) $ with kernel $\mathcal{A}( 1) $.\ \ This is just the abelian
group that is not considered further here.\ \ \ The faithful representation
with $\lambda \neq 0$ requires $p=0$, and therefore has the trivial
little group $\mathcal{K}\mbox{}^{\circ}\simeq \text{\boldmath $e$}\simeq
\{\Upsilon ( 0,0,0) \}$.\ \ The stabilizer is $\mathcal{G}\mbox{}^{\circ}\simeq
\mathcal{A}( n+1) $.\ \ The orbits are 
\begin{equation}
\mathbb{O}_{\lambda }=\left\{ {\widehat{\varsigma }}_{\Upsilon ( 0,q,0)
}[ \xi _{\alpha ,\lambda }] |q\in \mathbb{R}^{n}\right\} =\left\{
\xi _{q,\lambda }|q\in \mathbb{R}^{n}\right\} ,\ \ \lambda \in \mathbb{R}\backslash
\left\{ 0\right\} .
\end{equation}

All representations in the orbit are equivalent for the determination
of the semidirect product unitary irreducible representations. A
convenient representative of the equivalence class is $\xi _{0,\lambda
}$.\ \ The unitary representations $\sigma $ of the trivial little
group are trivial and therefore the representations of the stabilizer
are just $\varrho \mbox{}^{\circ}=\xi _{0,\lambda }$.\ \ The Hilbert
space ${\text{\boldmath $\mathrm{H}$}}^{\sigma }$ is\ \ also trivial
and therefore the Hilbert space of the stabilizer is ${\text{\boldmath
$\mathrm{H}$}}^{\varrho \mbox{}^{\circ}}={\text{\boldmath $\mathrm{H}$}}^{\sigma
}\otimes {\text{\boldmath $\mathrm{H}$}}^{\xi }\simeq \mathbb{C}.$
\subsubsection{Mackey induction}

\noindent The final step is to apply the Mackey induction theorem
to determine the faithful unitary irreducible representations of
the full $\mathcal{H}( n) $ group. The induction requires the definition
of the symmetric space
\begin{equation}
\mathbb{K}=\mathcal{G}/\mathcal{G}\mbox{}^{\circ}=\mathcal{H}( n)
/\mathcal{A}( n+1) \simeq \mathcal{A}( n) \simeq \mathbb{R}^{n},
\end{equation}

\noindent with the natural projection $\pi $ and a section $\Theta
$\ \ \ 
\begin{equation}
\begin{array}{l}
 \pi :\mathcal{H}( n)  \rightarrow \mathbb{K}:\Upsilon ( p,q,\iota
) \mapsto {\mathrm{k}}_{q}, \\
 \Theta :\mathbb{K}\rightarrow \mathcal{H}( n) :{\mathrm{k}}_{q}\mapsto
\Theta ( {\mathrm{k}}_{q}) =\Upsilon ( 0,q,0) .
\end{array}
\end{equation}

\noindent These satisfy $\pi ( \Theta ( {\mathrm{a}}_{q}) ) ={\mathrm{a}}_{q}$
and so $\pi \circ \Theta ={\mathrm{Id}}_{\mathbb{K}}$ as required.
Using (2), an element of the Weyl-Heisenberg group\ \ $\mathcal{H}(
n) $ can be written as, 
\begin{equation}
\Upsilon ( p,q,\iota ) =\Upsilon ( 0,q,0) \Upsilon ( p,0,\iota +\frac{1}{2}p\cdot
q) .
\end{equation}

\noindent The cosets are therefore defined by 
\begin{equation}
\begin{array}{ll}
 {\mathrm{k}}_{q} & =\left\{ \Upsilon ( 0,q,0) \Upsilon ( p,0,\iota
+\frac{1}{2}p\cdot q) |p\in \mathbb{R}^{n},\iota \in \mathbb{R}\right\}
\\
  & =\left\{ \Upsilon ( 0, q, 0) \mathcal{A}( n + 1) \right\} 
\end{array}
\end{equation}

\noindent Note that
\begin{equation}
\ \ \Upsilon ( p,q,\iota ) {\mathrm{k}}_{x} ={\mathrm{k}}_{x+q},\ \ \ \ x\in
\mathbb{R}^{n}.%
\label{PH: action of group on coset}
\end{equation}

\noindent The Mackey induced representation theorem can now be applied
straightforwardly.\ \ \ First, the Hilbert space is 
\begin{equation}
{\text{\boldmath $\mathrm{H}$}}^{\varrho }={\text{\boldmath $L$}}^{2}(
\mathbb{K},{\text{\boldmath $\mathrm{H}$}}^{\varrho \mbox{}^{\circ}})
\simeq {\text{\boldmath $L$}}^{2}( \mathbb{R}^{n},\mathbb{C}) .
\end{equation}

\noindent Next the Mackey induction Theorem 8 yields
\begin{equation}
\psi ^{\prime }( {\mathrm{k}}_{x}) =\left( \varrho ( \Upsilon (
p,q,\iota ) )  \psi \right) \left( {\Upsilon ( p,q,\iota ) }^{-1}{\mathrm{k}}_{x}\right)
=\varrho \mbox{}^{\circ}( \Upsilon ( a \mbox{}^{\circ},0,\iota \mbox{}^{\circ})
) \psi ( {\mathrm{k}}_{x-q}) 
\end{equation}

\noindent Using the Weyl-Heisenberg group product (2),
\begin{equation}
\begin{array}{ll}
 \Upsilon ( \mathit{p\mbox{}^{\circ}},\mathit{q\mbox{}^{\circ}},\iota
\mbox{}^{\circ})  & ={\Theta ( {\mathrm{k}}_{x}) }^{-1}\Upsilon
( p,q,\iota )  \Theta ( {\Upsilon ( p,q,\iota ) }^{-1}{\mathrm{k}}_{x})
\\
  & =\Upsilon ( 0, -x, 0) \Upsilon ( p, q, \iota ) \Upsilon ( 0,
x - q, 0)  \\
  & =\Upsilon ( p,0,\iota +p\cdot \left( x-\frac{1}{2}q\right) )
.
\end{array}
\end{equation}

\noindent  We lighten notation using the isomorphism ${\mathrm{k}}_{x}\mapsto
x$.\ \ The induced representation theorem then yields
\begin{equation}
\begin{array}{ll}
 \psi ^{\prime }( x)  & =\xi _{0,\lambda }(\Upsilon ( p,0,\iota
+x\cdot p-\frac{1}{2}p\cdot q) \psi ( x-q)  \\
  & =e^{i \lambda ( \iota +x\cdot p-\frac{1}{2}p\cdot q) }\psi (
x-q) .
\end{array}%
\label{PH: WH UIR}
\end{equation}

\noindent Using Taylor expansion, we can write
\begin{equation}
\psi ( x-q) = e^{-q^{i} \frac{\partial }{\partial  {x}^{i}}}\psi
( x) .
\end{equation}

\noindent The\ \ Baker Campbell-Hausdorff formula [20] enables us
to combine the exponentials
\begin{equation}
\psi ^{\prime }( x) =e^{i \left(  \lambda  \iota +\lambda  p^{i}
x_{i}+ q^{i}i\frac{\partial }{\partial x^{i}}\right) }\psi ( x)
=e^{i \left( \iota \widehat{I}+p^{i}{\widehat{Q}}_{i} +q^{i}{\widehat{P}}_{i}
\right) }\psi ( x) .%
\label{PH: WH general nondegenerate representations}
\end{equation}

The representation of the algebra is therefore
\begin{equation}
\widehat{I}\psi ( x) =\lambda  \psi ( x)  ,\ \ \ {\widehat{Q}}_{i}\psi (
x) =\lambda  x_{i}\psi ( x) ,\ \ \ {\widehat{P}}_{i}\psi ( x) =i \frac{\partial
}{\partial  x^{i}}\psi ( x) ,%
\label{PH: WH general algebra}
\end{equation}

\noindent that satisfies the Heisenberg commutation relations (1).

This analysis can also be carried out choosing $\Upsilon ( 0,q,\iota
) \in \mathcal{A}( n+1) $ to be the elements of the normal subgroup
and this yields the representation with ${\widehat{P}}_{i}$ diagonal.\ \ 
\subsection{Unitary irreducible representations of $\mathcal{H}\overline{\mathcal{S}p}(
2n) $}\label{MG: Section uir hsp}

We consider next the unitary irreducible representations of the
$\mathcal{H}\overline{\mathcal{S}p}( 2n) $ group\ \ \ \ 
\begin{equation}
\mathcal{H}\overline{\mathcal{S}p}( 2n) \simeq \mathcal{S}p( 2n)
\otimes _{s}\mathcal{H}( n) .
\end{equation}

As $\mathcal{H}\overline{\mathcal{S}p}( 2n) $\ \ is the central
extension of $\mathcal{I}\mathcal{S}p( 2n) $, the projective representations
of $\mathcal{I}\mathcal{S}p( 2n) $ are equivalent to the ordinary
unitary representations of $\mathcal{H}\overline{\mathcal{S}p}(
2n) $. 

The unitary irreducible representations of $\mathcal{H}\overline{\mathcal{S}p}(
2n) $ may be determined using Mackey Theorem 9 for the nonabelian
normal subgroup case.\ \ \ The faithful unitary representations
of the Weyl-Heisenberg group are given in the previous section (105).
The next step in applying the Mackey's theorem is to determine the
$\rho $ representation of the stabilizer $\mathcal{G}\mbox{}^{\circ}\subset
\mathcal{H}\overline{\mathcal{S}p}( 2n) $.
\subsubsection{Stabilizer and $\rho $ representation}

The representation $\rho $ of the stabilizer $\mathcal{G}\mbox{}^{\circ}$
acts on the Hilbert space ${\text{\boldmath $\mathrm{H}$}}^{\xi
}$ and therefore the hermitian representations $\rho ^{\prime }$
of the algebra of the stabilizer must be realized in the enveloping
algebra of the Weyl-Heisenberg group. The $\rho $ representation
restricted to the Weyl-Heisenberg group are given by $\rho |_{\mathcal{H}(
n) }=\xi $ where $\xi $ are the unitary irreducible representations
of the Weyl Heisenberg group.\ \ The faithful representations $\xi
$ are given in (105). 

The unitary representation $\rho $ acts on ${\text{\boldmath $\mathrm{H}$}}^{\xi
}\simeq {\text{\boldmath $L$}}^{2}( \mathbb{R}^{n},\mathbb{C}) $
such that
\begin{equation}
\rho ( \Omega \mbox{}^{\circ}) \xi ( \Upsilon ( z,\iota ) ) {\rho
( \Omega \mbox{}^{\circ}) }^{-1}= \xi ( \varsigma _{\Omega \mbox{}^{\circ}}\Upsilon
( z,\iota ) ) ,\ \ \Omega \mbox{}^{\circ}\in \mathcal{G}\mbox{}^{\circ}.
\end{equation}

\noindent The representation $\rho$ factors into\ \ 
\begin{equation}
\rho ( \Omega \mbox{}^{\circ}( \delta ,\Sigma ,w,r) ) = \xi ( \Upsilon
( w,r) ) \rho ( \Sigma ) ,
\end{equation}

\noindent where again for notational brevity\ \ $\Sigma \equiv \Omega
( 1,\Sigma ,0,0) $.

We already have characterized the inner automorphisms. The automorphisms
corresponding factor as
\begin{equation}
\begin{array}{l}
 \begin{array}{l}
 \xi ( \Upsilon ( w,r) ) \xi ( \Upsilon ( z,\iota ) ) {\xi ( \Upsilon
( w,r) ) }^{-1}= \xi ( \varsigma _{\Upsilon ( w,r) }\Upsilon ( z,\iota
) ) , \\
 \rho ( \Sigma ) \xi ( \Upsilon ( z,\iota ) ) {\rho ( \Sigma ) }^{-1}=
\xi ( \varsigma _{\Omega ( \Sigma ) }\Upsilon ( z,\iota ) ) =\xi
( \Upsilon ( \pi ( \Sigma ) z,\iota ) ) .
\end{array}
\end{array}
\end{equation}

\noindent where $\Sigma \in \overline{\mathcal{S}p}( 2n) $ and $\pi
:\overline{\mathcal{S}p}( 2n) \rightarrow \mathcal{S}p( 2n) $. 

The inner automorphisms are already characterized as we know the
unitary irreducible representations $\xi $. Consider next the representation
$\rho ( \Sigma ) $ of the symplectic group $\overline{\mathcal{S}p}(
2n) $.\ \ The hermitian representation of the symplectic generators
is
\begin{equation}
\begin{array}{l}
 \begin{array}{l}
 {\widehat{A}}_{i,j}=\rho ^{\prime }( A_{i,j} ) =\lambda  {\widehat{Q}}_{i}{\widehat{P}}_{j},
\\
 {\widehat{B}}_{i,j}=\rho ^{\prime }( B_{i,j}) =\lambda {\widehat{Q}}_{i}{\widehat{Q}}_{j},
\\
 {\widehat{C}}_{i,j}=\rho ^{\prime }( C_{i,j}) =\lambda  {\widehat{P}}_{i}{\widehat{P}}_{j}.
\end{array}
\end{array}
\end{equation}

Clearly ${\widehat{B}}_{i,j}={\widehat{B}}_{j,i}$ and ${\widehat{C}}_{i,j}={\widehat{C}}_{j,i}$.\ \ Then,
using the Heisenberg commutation relations (1), this defines a hermitian
realization of the Lie algebra of the automorphism group acting
on the Hilbert space ${\text{\boldmath $\mathrm{H}$}}^{\xi }\simeq
{\text{\boldmath $L$}}^{2}( \mathbb{R}^{n},\mathbb{C}) $.
\begin{gather}
 \begin{array}{l}
 \begin{array}{l}
 \left[ {\widehat{A}}_{i,j},{\widehat{A}}_{k,l}\right]  =i( \delta _{i,l}{\widehat{A}}_{j,k}-\delta
_{j,k}{\widehat{A}}_{i,l}) , \\
 \left[ {\widehat{A}}_{i,j},{\widehat{B}}_{k,l}\right]  =i( \delta _{j,k}{\widehat{B}}_{i,l}+\delta
_{j,l}{\widehat{B}}_{i,k}) , \\
 \left[ {\widehat{A}}_{i,j},{\widehat{C}}_{k,l}\right]  =-i( \delta _{i,k}{\widehat{C}}_{j,l}+\delta
_{i,l}{\widehat{C}}_{k,j}) , \\
 \left[ {\widehat{B}}_{i,j},{\widehat{C}}_{k}\right]  =i( \delta _{i,k}{\widehat{A}}_{j,l}
+\delta _{i,l}{\widehat{A}}_{j,k} +\delta _{j,k}{\widehat{A}}_{i,l} +\delta
_{j,l}{\widehat{A}}_{i,k} ) ,
\end{array}
\end{array}
\\\begin{array}{ll}
 \left[ {\widehat{A}}_{i,j},{\widehat{Q}}_{k}\right] = i \delta _{j,k}{\widehat{Q}}_{i},
& \left[ {\widehat{C}}_{i,j},{\widehat{Q}}_{k}\right] =i( \delta _{j,k}{\widehat{P}}_{i}+
\delta _{i,k}{\widehat{P}}_{j}) , \\
 \left[ {\widehat{A}}_{i,j},{\widehat{P}}_{k}\right] = -i \delta _{i,k}{\widehat{P}}_{j},
& \left[ {\widehat{B}}_{i,j},{\widehat{P}}_{k}\right] =i(  \delta _{j,k}{\widehat{Q}}_{i}+\delta
_{i,k}{\widehat{Q}}_{j}) , \\
 \left[ {\widehat{P}}_{i},{\widehat{Q}}_{j}\right] =i \delta _{i,j}\widehat{I}.
&  
\end{array}
\end{gather}

Therefore, there exists a $\rho ^{\prime }$ representation for the
entire algebra of $\mathcal{H}\overline{\mathcal{S}p}( 2n) $ and
therefore the stabilizer is the group itself, $\mathcal{G}\mbox{}^{\circ}\simeq
\mathcal{H}\overline{\mathcal{S}p}( 2n) $. This explicate construction
of the algebra shows that the representation $\rho ( \Sigma ) $
exists.\ \ Consequently, the Mackey induction theorem is not required.

The $\rho ( \Sigma ) $ representation is precisely (up to an overall
phase) the metaplectic representation originally studied by Weil
\cite{weil}, \cite{folland}. We can construct this explicitly using
the factorization of the symplectic group (86).\ \ We can consider
each of the factors separately as 
\begin{equation}
\rho (\Sigma ( \epsilon ,\alpha ,\beta ,\gamma ) =\rho ( \Sigma
^{-}( \gamma ) ) \rho ( \Sigma \mbox{}^{\circ}( \alpha ) ) \rho
( \Sigma ^{+}( \beta ) ) \rho ( \zeta ^{ \epsilon }) ,
\end{equation}

\noindent and each of these factors can be applied separately to
determine the $\rho $ representation.\ \ The unitary representations
of $\Sigma ( \beta ) \in \mathcal{A}( m) $, $m=\frac{n( n+1) }{2}$
in a basis with ${\widehat{Q}}_{i}$ diagonal are
\begin{equation}
\begin{array}{l}
 \left. \rho ( \Sigma ^{+}( \beta ) ) |\psi _{\lambda }( x) \right\rangle
=e^{i \alpha ^{i,j}{\widehat{B}}_{i,j}} \overset{ }{\left. |\psi _{\lambda
}( x) \right\rangle  }=e^{\frac{i}{\lambda }\beta ^{i,j}x_{i}x_{j}}
\overset{ }{\left. |\psi _{\lambda }( x) \right\rangle  }
\end{array}.%
\label{MG: metaplectic beta}
\end{equation}

The representations of the elements of the unitary group $\Sigma
( \alpha ) \in \mathcal{U}( n) $ are
\begin{equation}
\begin{array}{l}
 \left. \rho ( \Sigma \mbox{}^{\circ}( \alpha ) ) |\psi _{\lambda
}( x) \right\rangle  ={\overset{ }{|\det  A |}}^{-\frac{1}{2}} \overset{
}{\left. |\psi _{\lambda }( A^{-1}x) \right\rangle  }
\end{array}.
\end{equation}

\noindent The symplectic matrix exchanges the $p$ and $q$ degrees
of freedom, $\varsigma _{\zeta }\Upsilon ( p,q,\iota ) =\Upsilon
( q,-p,\iota ) $. As is well known, the unitary representation of
this is the Fourier transform, $\rho ( \zeta ) =\text{\boldmath
$f$}$ where 
\begin{equation}
\left. \left. \rho ( \Upsilon ( p,q,\iota ) ) \text{\boldmath $f$}
|\psi _{\lambda }( x) \right\rangle  =\text{\boldmath $f$} \rho
( \Upsilon ( q,-p,\iota ) ) |\psi _{\lambda }( x) \right\rangle
,
\end{equation}

\noindent where the Fourier transform is defined as usual by 
\begin{equation}
\widetilde{\psi }( y) =\text{\boldmath $f$}\psi ( x) = {\left( 2 \pi
i\right) }^{-\frac{n}{2}}\int e^{-i x \cdot y}\psi ( x) d^{n}x,
\end{equation}

\noindent and where 
\begin{equation}
{\widehat{Q}}_{i}\overset{ }{\left. |\psi _{\lambda }( x) \right\rangle
}=\lambda  x_{i}\overset{ }{\left. |\psi _{\lambda }( x) \right\rangle
},\ \ {\widehat{P}}_{i}\overset{ }{\left. |{\widetilde{\psi }}_{\lambda
}( y) \right\rangle  }=y_{i}\overset{ }{\left. |\widetilde{\psi _{\lambda
}}( y) \right\rangle  }.
\end{equation}

Finally, the $\rho ( \Sigma ^{+}( \beta ) ) $ representation can
be computed using (80) in a basis with ${\widehat{Q}}_{i}$ diagonal
giving 
\begin{equation}
\rho ( \Sigma ^{-}( \gamma ) ) \overset{ }{\left. |\psi _{\lambda
}( x) \right\rangle  }=\text{\boldmath $f$} \rho ( \Sigma ^{+}(
-\gamma ) ) {\text{\boldmath $f$}}^{-1}|\psi _{\lambda }( x) ,
\end{equation}

\noindent and the $\rho ( \Sigma ^{+}( -\gamma ) ) $ is given by
(116).\ \ Putting all of these together gives the representation
$\rho ( \Sigma ) $ up to a phase.\ \ While one would expect the
phase to be $m\in \mathbb{Z}$ dependent, it actually only is two
valued $\pm 1\in \mathbb{Z}_{2}$.\ \ The unitary representations
of the double cover metaplectic group $\mathcal{M}p( 2n) $ are also
a representation of $\overline{\mathcal{S}p}( 2n) $ due to the homomorphism
(141).\ \ 

Of course, all of these calculations could also be done in a basis
with ${\widehat{P}}_{i}$ diagonal.

As the stabilizer is the full group, Mackey induction is not required
and the unitary irreducible representations $\upsilon $ of $\mathcal{H}\overline{\mathcal{S}p}(
2n) $ are given by
\begin{equation}
\left. \left. \upsilon ( \Omega ( 1, \Sigma ,z,\iota ) ) |\psi (
x) \right\rangle  =\sigma ( \Sigma ) \otimes \xi ( \Upsilon ( z,\iota
) ) \rho ( \Sigma ) |\psi ( x) \right\rangle  %
\label{MG: HSp unitary irreps}
\end{equation}

\noindent where $\sigma $ are ordinary unitary irreducible representations
of $\overline{\mathcal{S}p}( 2n) $, $\rho $ are the metaplectic
representation of $\overline{\mathcal{S}p}( 2n) $ given above and
$\xi $ are the unitary irreducible representations of $\mathcal{H}(
n) $ given in Section 3.1. 

The ordinary unitary representations of the symplectic group have
been partially characterized [8-9]. A complete set of unitary irreducible
representations of the covering group $\overline{\mathcal{S}p}(
2n) $ appears to be an open problem.
\section{Summary}

We have determined the projective representations of the inhomogeneous
symplectic group. This is the maximal symmetry whose projective
representations transform physical states such that the Heisenberg
commutation relations are valid in all of the transformed states.

The inhomogeneous symplectic symmetry is well known from classical
mechanics.\ \ It acts on classical phase space with position and
momentum degrees of freedom. The projective representations that
define the quantum symmetry require its central extension which
introduces the non-abelian structure of the Weyl-Heisenberg group,
$\mathcal{I}\widecheck{\mathcal{S}p}( 2n) \simeq \overline{\mathcal{S}p}(
2n) \otimes _{s}\mathcal{H}( n) $.\ \ The non-abelian structure
is a direct result of the fact that transition probabilities are
the square of the norm of physical states. Consequently, the physical
states are defined up to a phase and the action of a symmetry group
is given by the projective representations.\ \ This is the underlying
reason for the non-abelian structure, or {\itshape quantization}.\ \ Any
symmetry of quantum mechanics that preserves the position, momentum
Heisenberg commutation relations must be a subgroup of this maximal
symmetry. 

On the other hand, we now understand special relativistic quantum
mechanics as the projective representations of the inhomogeneous
Lorentz group \cite{wigner},\cite{Weinberg1}.\ \ The central extension
of this group does not admit an algebraic extension. For the connected
component, the central extension is therefore the cover that we
call the Poincar\'e group which for $n=3$ is $\mathcal{P}=\mathcal{S}\mathcal{L}(
2,\mathbb{C}) \otimes _{s}\mathcal{A}( 4) $.\footnote{The full inhomogeneous
group is given in terms of the orthogonal group $\mathcal{O}( 1,n)
$ that has 4 disconnected components. The discrete $\mathbb{Z}_{2,2}$
symmetry is P, T and PT symmetry.\ \ Its central extension is not
unique and it gives rise to the $\mathcal{P}in$ group ambiguity.\ \ On
the other hand the $\mathcal{S}\mathcal{O}( 1,n) $ group has 2 components
but does have a unique central extension that is the $\mathcal{S}pin$
group.\ \ The discrete $\mathbb{Z}_{2}$ symmetry is the PT symmetry.
}\ \ Special relativistic quantum mechanics is formulated in terms
of the unitary representations of the Poincar\'e group.\ \ There
is however, no mention of the Weyl-Heisenberg group which plays
a fundamental role in the original formulation of quantum mechanics.

Symmetry is one of the most fundamental concepts of physics. We
have the case where we have a quantum symmetry for the Weyl-Heisenberg
of quantum mechanics that is the projective representations of a
classical symmetry on phase space. On the other hand, the quantum
symmetry for the Minkowski metric of special relativity is given
in terms of a classical symmetry on position-time space, that is,
spacetime.\ \ 

Quantum mechanics and special relativity have at best, an uneasy
marriage.\ \ Perhaps it is due to this underlying disparity in the
most basic symmetries of these theories. The standard approach is
to ignore the quantum symmetry described in this paper and formulate
special relativistic quantum mechanics as the projective representations
of the inhomogeneous group. 

If we truly are to bring together quantum mechanics and special
relativity, we must first reconcile these basic symmetries and find
a symmetry that encompasses both.\ \ This can be done in a remarkably
straightforward manner and results in a theory that, in a physical
limit, results in the usual formulation of special relativistic
quantum mechanics.\ \ But, before the limit is taken, it points
to a theory incorporating both symmetries that may give further
understanding of the unification of quantum mechanics and relativity
\cite{Low5},\cite{Low6},\cite{Low14}.\ \ In this theory, physics
takes place in extended phase space and there is no invariant global
projection that gives physics in position-time space (i.e. space-time).\ \ \ Generally,
local observers with general non-inertial trajectories construct
different space-times as subspaces of extended phase space. The
usual Lorentz symmetry continues to hold exactly for inertial trajectories
but is generalized in a remarkable manner for non-inertial trajectories.\ \ \ \ 
\section{Appendix A: Key Theorems}
In this appendix we review a set of definitions and theorems that
are fundamental for the application of symmetry groups in quantum
mechanics.\ \ We state the theorems only and refer the reader to
the cited literature for full proofs.
\begin{definition}

A group $\mathcal{G}$ is a semidirect product if it has a subgroup
$\mathcal{K}$ (referred to as the homogeneous subgroup) and a normal
subgroup $\mathcal{N}$ such that $\mathcal{K}\cap \mathcal{N}=\text{\boldmath
$e$}$ and $\mathcal{G}\simeq \mathcal{N} \mathcal{K}$.\ \ \ Our
notation for a semidirect product is $\mathcal{G}\simeq \mathcal{K}\otimes
_{s}\mathcal{N}$\footnote{Our notation follows [17].\ \ Another
notation commonly used is $\mathcal{N}\rtimes \mathcal{K}$.\ \ It
is just notation; the definition remains the same for both notations.
}\cite{sternberg}. \label{PH: def: semidirect product}
\end{definition}
It follows directly that a semidirect product is right associative
in the sense that $\mathcal{D}\simeq (\mathcal{A}\otimes _{s}\mathcal{B})\otimes
_{s}\mathcal{C}$ implies that $\mathcal{D}\simeq \mathcal{A}\otimes
_{s}(\mathcal{B}\otimes _{s}\mathcal{C})$ and so brackets can be
removed. However $\mathcal{D}\simeq \mathcal{A}\otimes _{s}(\mathcal{B}\otimes
_{s}\mathcal{C})$ does not necessarily imply $\mathcal{D}\simeq
(\mathcal{A}\otimes _{s}\mathcal{B})\otimes _{s}\mathcal{C}$ as
$\mathcal{B}$ is not necessarily a normal subgroup of $\mathcal{A}$.
\begin{definition}

An algebraic central extension of a Lie algebra $g$ is the Lie algebra
$\widecheck{g}$ that satisfies the following short exact sequence where
$z$ is the maximal abelian algebra that is central in $\widecheck{g}$,\label{PH:
def: alg central extension}
\end{definition}
\begin{equation}
\text{\boldmath $0$}\rightarrow z\rightarrow \widecheck{g}\rightarrow
g\rightarrow \text{\boldmath $0$} .
\end{equation}

\noindent where $\text{\boldmath $0$}$ is the trivial algebra. Suppose
$\{X_{a}\}$ is a basis of the Lie algebra $g$ with commutation relations
$[X_{a},X_{b}]=c_{a,b}^{c}X_{c}$, $a,b=1,...r$.\ \ Then an algebraic
central extension is a maximal set of central abelian generators
$\{A_{\alpha }\}$, where $\alpha ,\beta ,... =1,..m$,\ \ such that
\begin{equation}
\left[ A_{\alpha },A_{\beta }\right] =0,\ \ \ \ \left[ X_{a},A_{\alpha
}\right] =0,\ \ \ \ \left[ X_{a},X_{b}\right] =c_{a,b}^{c}X_{c}+c_{a,b}^{\alpha
}A_{\alpha }.
\end{equation}

\noindent The basis $\{X_{a},A_{\alpha }\}$ of the centrally extended
Lie algebra must also satisfy the Jacobi identities. The Jacobi
identities\ \ constrain the admissible central extensions of the
algebra. The choice\ \ $X_{a}\mapsto X_{a}+A_{a}$ will always satisfy
these relations and this trivial case is excluded.\ \ The algebra
$\widecheck{g}$ constructed in this manner is equivalent to the central
extension of $g$ given in Definition 2.
\begin{definition}

The central extension of a connected Lie group $\mathrm{\mathcal{G}}$
is the\ \ Lie group $\widecheck{\mathcal{G}}$ that satisfies the following
short exact sequence where $\mathrm{\mathcal{Z}}$ is a maximal abelian
group that is central in $\widecheck{\mathcal{G}}$
\end{definition}
\begin{equation}
\text{\boldmath $e$}\rightarrow \mathcal{Z}\rightarrow \widecheck{\mathcal{G}}\overset{\pi
}{\rightarrow }\mathcal{G}\rightarrow \text{\boldmath $e$} .
\end{equation}

The abelian group $\mathcal{Z}$ may always be written as the direct
product $\mathcal{Z}\simeq \mathcal{A}( m) \otimes \mathbb{A}$ of
a connected continuous abelian Lie group $\mathcal{A}( m) \simeq
(\mathbb{R}^{m},+)$ and a discrete abelian group $\mathbb{A}$ that
may have a finite or countable dimension [10].

The exact sequence may be decomposed into an exact sequence for
the {\itshape topological} central extension and the {\itshape algebraic}
central extension,
\begin{equation}
\text{\boldmath $e$}\rightarrow \mathbb{A}\rightarrow \overline{\mathcal{G}}\overset{\pi
\mbox{}^{\circ}}{\rightarrow }\mathcal{G}\rightarrow \text{\boldmath
$e$} \text{\boldmath $,$}\text{\boldmath $\ \ $}\text{\boldmath
$e$}\rightarrow \mathcal{A}( m) \rightarrow \widecheck{\mathcal{G}}\overset{\widetilde{\pi
}}{\rightarrow }\overline{\mathcal{G}}\rightarrow \text{\boldmath
$e$}.
\end{equation}

\noindent where $\pi =\pi \mbox{}^{\circ}\circ \widetilde{\pi }$. The
first exact sequence defines the universal cover where $\mathbb{A}\simeq
\ker  \pi \mbox{}^{\circ}$ is the fundamental homotopy group. All
of the groups is in the second sequence are simply connected and
therefore may be defined by the exponential map of the central extension
of the algebra given by Definition 2. In other words, the full central
extension may be computed by determining the universal covering
group of the algebraic central extension.
\begin{definition}

A ray $\Psi $ is the equivalence class of states $|\psi _{\gamma
}\rangle $ that are elements of a Hilbert space $\text{\boldmath
$\mathrm{H}$}$ up to a phase, 
\end{definition}
\begin{equation}
\left. \left. \Psi =\left\{ e^{i \omega }\left| \psi \right. \right.
\right\rangle   |\omega \in \mathbb{R}\right\} ,\ \ \ \left. \left|
\psi \right. \right\rangle  \in H.
\end{equation}

\noindent Note that the physical probabilities that are the square
of the modulus depend only on the ray
\[
|\left( \Psi _{\beta },\Psi _{\alpha }\right) |^{2}={\left| \left\langle
\psi _{\beta }|\psi _{\alpha }\right\rangle  \right| }^{2}
\]

\noindent for all $|\psi _{\gamma }\rangle \in \Psi $. For this
reason, physical states in quantum mechanics are defined to be rays
rather than states in the Hilbert space
\begin{definition}

A projective representation $\varrho $ of a symmetry group $\mathcal{G}$$\text{}$
is the maximal representation such that\ \ for $|{\widetilde{\psi }}_{\gamma
}\rangle =\varrho ( g) |\psi _{\gamma }\rangle $, the modulus is
invariant\ \ ${|\langle {\widetilde{\psi }}_{\beta }|{\widetilde{\psi }}_{\alpha
}\rangle |}^{2}={|\langle \psi _{\beta }|\psi _{\alpha }\rangle
|}^{2}$ for all $|\psi _{\gamma }\rangle , |{\widetilde{\psi }}_{\gamma
}\rangle \in \Psi $.\label{PH: def: proj representation}
\end{definition}
\begin{theorem}

{\bfseries (Wigner, Weinberg):} Any projective representation of
a Lie symmetry group $\mathcal{G}$ on a separable Hilbert space
is equivalent to a representation that is either linear and unitary
or anti-linear and anti-unitary. Furthermore, if $\mathcal{G}$ is
connected, the projective representations are equivalent to a representation
that is linear and unitary [1],[11].\label{PH: theorem: Wigner unitary
projective}
\end{theorem}

This is the generalization of the well known theorem that the ordinary
representation of any compact group is equivalent to a representation
that is unitary.\ \ For a projective representation, the phase degrees
of freedom of the central extension enables the equivalent linear
unitary or anti-linear anti-unitary representation to be constructed
for this much more general class of Lie groups that admit representations
on separable Hilbert spaces.\ \ (A proof of the theorem is given
in Appendix A of Chapter 2 of \cite{Weinberg1}).\ \ The set of groups
that this theorem applies to include all the groups that are studied
in this paper.
\begin{theorem}

{\bfseries (Bargmann, Mackey)} The projective representations of
a connected Lie group $ \mathcal{G}$ are equivalent to the ordinary
unitary representations of its central extension $\widecheck{\mathcal{G}}$
\cite{bargmann}, \cite{mackey2}.\label{PH: theorem: proj rep is
unitary CE}
\end{theorem}

Theorem 2 states that are all projective representations of a connected
Lie group are equivalent to a projective representation that is
unitary. A phase is the unitary representation of a central abelian
subgroup. Therefore, the maximal representation is given in terms
of the central extension of the group. 
\begin{theorem}

Let $\mathcal{G}$,$\mathcal{H}$ be Lie groups and $\pi :\mathcal{G}\rightarrow
\mathcal{H}$ be a homomorphism. Then, for every unitary representation
$\widetilde{\varrho }$ of $\mathcal{H}$ there exists a degenerate unitary
representation $\varrho $ of $\mathcal{G}$ defined by $\varrho =\widetilde{\varrho
}\circ \pi $. Conversely, for every degenerate unitary representation
of a Lie group $\mathcal{G}$ there exists a Lie subgroup $\mathcal{H}$
and a homomorphism $\pi :\mathcal{G}\rightarrow \mathcal{H}$ where
$\ker ( \pi ) \neq \text{\boldmath $e$}$ such that $\varrho =\widetilde{\varrho
}\circ \pi $\ \ where $\widetilde{\varrho }$ is a unitary representation
of $\mathcal{H}$.\label{PH: theorem: degenerate reps}
\end{theorem}

Noting that a representation is a homomorphism, This theorem follows
straightforwardly from the properties of homomorphisms. As a consequence,
the set of degenerate representations of a group is characterized
by its set of normal subgroups. A {\itshape faithful} representation
is the case that the representation is an isomorphism.
\begin{theorem}

{\bfseries (Levi) }Any simply connected\ \ Lie group is equivalent
to the semidirect product of a semisimple group and a maximal solvable
normal subgroup{\bfseries  }\cite{barut}\label{PH: theorem: Levi}
\end{theorem}

As the central extension of any connected group is simply connected,
the problem of computing the projective representations of a group
always can be reduced to computing the unitary irreducible representations
of a semidirect product group with a semisimple homogeneous group
and a solvable normal subgroup. The unitary irreducible representations
of the semisimple groups are known and the solvable groups that
we are interested in turn out to be the semidirect product of abelian
groups.
\begin{theorem}

Any semidirect product group $\mathcal{G}\simeq \mathcal{K}\otimes
_{s}\mathcal{N}$ is a subgroup of a group homomorphic to the group
of automorphisms of $\mathcal{N}$ [13].\label{PH: theorem: automorphisms
semid-direct}
\end{theorem}

The proof follows directly from the definition of the semidirect
product and an automorphism group.
\begin{theorem}

The automorphism group of a simply connected group is isomorphic
to the automorphism group of its Lie algebra.\cite{barut}\label{PH:
theorem: automorphisms simply connected}
\end{theorem}
\subsection{Mackey theorems for the representations of semidirect
product groups}

The Mackey theorems are valid for a general class of topological
groups but we will only require the more restricted case $\mathcal{G}\simeq
\mathcal{K}\otimes _{s}\mathcal{N}$ where the group $\mathcal{G}$
and subgroups $\mathcal{K},\mathcal{N}$ are smooth Lie groups.\ \ The
central extension of any connected Lie group is simply connected
and therefore generally has the form of a semidirect product due
to Theorem 5 (Levi). Theorem 6 further constrains the possible homogeneous
groups $\mathcal{K}$ of the semidirect product given the normal
subgroup $\mathcal{N}$.

The first Mackey theorem is the induced representation theorem that
gives a method of constructing a unitary representation of a group
(that is not necessarily a semidirect product group) from a unitary
representation of a closed subgroup. The second theorem gives a
construction of certain representations of a certain subgroup of
a semidirect product group from which the complete set of unitary
irreducible representations of the group can be induced. This theorem
is valid for the general case where the normal subgroup $\mathcal{N}$
is a nonabelian group. In the special case where the normal subgroup
$\mathcal{N}$ is abelian, the theorem may be stated in a simpler
form.
\begin{theorem}

{\bfseries {\upshape (Mackey).}} {\upshape Induced representation
theorem.} Suppose that $\mathcal{G}$ is a Lie group and $\mathcal{H}$
is a Lie subgroup, $\mathcal{H}\subset \mathcal{G}$ such that $\mathbb{K}\simeq
\mathcal{G}/\mathcal{H}$ is a homogeneous space with a natural projection\ \ $\pi
:\mathcal{G}\rightarrow \mathbb{K}$, an invariant measure and a
canonical section {\upshape $\Theta :\mathbb{K}\rightarrow \mathcal{G}:\mathrm{k}\mapsto
g$} such that\ \ $\pi \circ \Theta \mathrm{=}{\mathrm{Id}}_{\mathbb{K}}$
where ${\mathrm{Id}}_{\mathbb{K}}$ is the identity map on $\mathrm{\mathbb{K}}$.
Let $\rho $ be a unitary representation of $\mathcal{H}$ on the
Hilbert space ${\text{\boldmath $\mathrm{H}$}}^{\rho }$:\label{PH:
theorem: Mackey induction theorem}
\end{theorem}
\[
\rho ( h) :{\text{\boldmath $\mathrm{H}$}}^{\rho }\rightarrow {\text{\boldmath
$\mathrm{H}$}}^{\rho }:\left. \left. \left| \varphi \right. \right\rangle
\mapsto \left| \widetilde{\varphi }\right. \right\rangle  =\left. \rho
( h) \left| \varphi \right. \right\rangle  ,\ \ h\in \mathcal{H}.
\]

\noindent Then a unitary representation $\mathrm{\varrho }$ of a
Lie group $\mathrm{\mathcal{G}}$ on the Hilbert space ${\text{\boldmath
$\mathrm{H}$}}^{\mathrm{\varrho }}$,
\[
\varrho ( g) :{\text{\boldmath $\mathrm{H}$}}^{\varrho }\rightarrow
{\text{\boldmath $\mathrm{H}$}}^{\varrho }:\left. \left. \left|
\psi \right. \right\rangle  \mapsto \left| \widetilde{\psi }\right.
\right\rangle  =\left. \varrho ( g) \left| \psi \right. \right\rangle
,\ \ g\in \mathcal{G},
\]

\noindent may be induced from the representation $\mathrm{\rho }$
of $\mathrm{\mathcal{H}}$\ \ by defining
\begin{equation}
\widetilde{\psi }( \mathrm{k}) =\left( \varrho ( g) \psi \right) \left(
\mathrm{k}\right) =\rho ( \mathit{g\mbox{}^{\circ}}) \psi ( g^{-1}\mathrm{k})
,\text{}\ \ \ \mathit{g\mbox{}^{\circ}}={\Theta ( \mathrm{k}) }^{-1}
g \Theta ( g^{-1}\mathrm{k}) ,
\end{equation}

\noindent where the Hilbert space on which the induced representation
$\mathrm{\varrho }$ acts is given by ${\text{\boldmath $\mathrm{H}$}}^{\varrho
}\simeq {\text{\boldmath $\mathrm{L}$}}^{2}( \mathbb{K},H^{\rho
}) $ [14], [13].

The proof is straightforward given that the section $\Theta $ exists
by showing first that $g \mbox{}^{\circ}\in \ker ( \pi ) \simeq
\mathcal{H}$ and therefore $\rho ( g \mbox{}^{\circ}) $ is well
defined.
\begin{definition}

({\itshape Little groups}): Let $\mathcal{G}=\mathcal{K}\otimes
_{s}\mathcal{N}$ be a semidirect product. Let $[\xi ]\in {\text{\boldmath
$U$}}_{\mathcal{N}}$ where\ \ ${\text{\boldmath $U$}}_{\mathcal{N}}$
denotes the unitary dual whose elements are equivalence classes
of unitary representations of $\mathcal{N}$ on a Hilbert space ${\text{\boldmath
$\mathrm{H}$}}^{\xi }$.\ \ Let $\rho $ be a unitary representation
of a subgroup $\mathcal{G}\mbox{}^{\circ}=\mathcal{K}\mbox{}^{\circ}\otimes
_{s}\mathcal{N}$ on the Hilbert space ${\text{\boldmath $\mathrm{H}$}}^{\xi
}$ such that $\rho _{|\mathcal{N}}=\xi $. The {\itshape little}
{\itshape groups} are the set of maximal subgroups $\mathcal{K}^{\circ
}$ such that $\rho $ exists on the corresponding {\itshape stabilizer}
$\mathcal{G}^{\circ }\simeq \mathcal{K}\mbox{}^{\circ}\otimes _{s}\mathcal{N}$
and satisfies the fixed point equation\label{PH: defn: little group}
\end{definition}
\begin{equation}
 {\widehat{\varsigma }}_{\rho ( k) }[ \xi ] =\left[ \xi \right]  , k\in
\mathcal{K}\mbox{}^{\circ}.%
\label{PH: Little group general equation}
\end{equation}

\noindent In this definition the dual automorphism is defined by
\begin{equation}
\begin{array}{ll}
 \left( {\widehat{\varsigma }}_{\rho ( g) }\xi \right) \left( h\right)
& =\rho ( g)  \rho ( h) {\rho ( g) }^{-1}=\rho ( g h g^{-1}) =\xi
(  \varsigma _{g}h) 
\end{array}%
\label{PH: automorphisms of little group}
\end{equation}

\noindent for all $g\in \mathcal{G}\mbox{}^{\circ}$ and $h\in \mathcal{N}$.\ \ The
equivalence classes of the unitary representations of $\mathcal{N}$
are defined by
\begin{equation}
 \left[ \xi \right] =\left\{ {\widehat{\varsigma }}_{\xi ( h) }\xi |h\in
\mathcal{N}\right\} .%
\label{PH: Abelian xi equivalence classess}
\end{equation}

\noindent A group $\mathcal{G}$ may have multiple little groups
${\mathcal{K}\mbox{}^{\circ}}_{\alpha }$ whose intersection is the
identity element only.\ \ We will generally leave the label $\alpha
$ implicit.
\begin{theorem}

{\bfseries {\upshape (Mackey).}} {\upshape Unitary irreducible representations
of semidirect products. }Suppose that we have a semidirect product
Lie group {\upshape $\mathcal{G}\simeq \mathcal{K}\otimes _{s}\mathcal{N}$},
where $\mathcal{K},\mathcal{N}$ are Lie subgroups. Let $\mathrm{\xi
}$ be the unitary irreducible representation of $\mathrm{\mathcal{N}}$
on the Hilbert space ${\text{\boldmath $\mathrm{H}$}}^{\xi }$.\ \ Let
$\mathcal{G}\mbox{}^{\circ}\simeq \mathcal{K}\mbox{}^{\circ}\otimes
_{s}\mathcal{N}$ be a maximal stabilizer on which there exists a
representation $\rho $ on ${\text{\boldmath $\mathrm{H}$}}^{\xi
}$ such that $\rho |_{\mathcal{N}}=\xi $.\ \ \ Let $\mathrm{\sigma
}$ be a unitary irreducible representation of $\mathrm{\mathcal{K}}\mbox{}^{\circ}$
on the Hilbert space ${\text{\boldmath $\mathrm{H}$}}^{\sigma }$.
Define the representation $ \varrho \mbox{}^{\circ}=\sigma \otimes
\rho $ that acts on the Hilbert space\ \ ${\text{\boldmath $\mathrm{H}$}}^{\varrho
\mbox{}^{\circ}}\simeq {\text{\boldmath $\mathrm{H}$}}^{\sigma }\otimes
{\text{\boldmath $\mathrm{H}$}}^{\xi }$. \label{PH: theorem: Mackey
semidirect product theorem}Determine the complete set of stabilizers
and representations $\rho $\ \ and little groups that satisfy these
properties, that we label by $\alpha $,$\{{(\mathcal{G}\mbox{}^{\circ},\varrho
\mbox{}^{\circ},{\text{\boldmath $\mathrm{H}$}}^{\varrho \mbox{}^{\circ}})}_{\alpha
}\}$.\ \ If for some member of this set $\mathrm{\mathcal{G}}\text{{\upshape
$ \mbox{}^{\circ}$ $ \simeq $ $ \mathcal{G}$}}$ then for this case
the representations\ \ are\ \ {\upshape $(\mathcal{G},\varrho ,{\text{\boldmath
$\mathrm{H}$}}^{\varrho })\mathrm{\simeq }(\mathcal{G}\mbox{}^{\circ},\varrho
\mbox{}^{\circ},{\text{\boldmath $\mathrm{H}$}}^{\varrho \mbox{}^{\circ}})$}.\ \ For
the cases where the stabilizer\ \ {\upshape $\mathcal{G}\mbox{}^{\circ}$}
is a proper subgroup of $\mathrm{\mathcal{G}}$ then the unitary
irreducible representations $(\mathcal{G},\varrho ,{\text{\boldmath
$\mathrm{H}$}}^{\varrho })$ are the representations induced (using
Theorem 8) by the representations {\upshape $(\mathcal{G}\mbox{}^{\circ},\varrho
\mbox{}^{\circ},{\text{\boldmath $\mathrm{H}$}}^{\varrho \mbox{}^{\circ}})$}
of the stabilizer subgroup. The complete set of unitary irreducible
representations is the union of the representations $\cup _{\alpha
}\{{(\mathcal{G},\varrho ,{\text{\boldmath $\mathrm{H}$}}^{\varrho
})}_{\alpha }\mathrm{\}}$ over the set of all the stabilizers and
corresponding little groups.
\end{theorem}

This major result and its proof are due to Mackey[14]. Our focus
in this paper is on applying this theorem.
\subsubsection{Abelian normal subgroup}

The theorem simplifies for special cases where the normal subgroup
$\mathcal{N}$ is an abelian group, $\mathcal{N}\simeq \mathcal{A}(
n) $.\ \ An abelian group has the property that its unitary irreducible
representations $\xi $ are the characters acting on the Hilbert
space ${\text{\boldmath $\mathrm{H}$}}^{\xi }\simeq \mathbb{C}$,
\begin{equation}
\xi ( a) \left| \phi \right\rangle  =e^{i a\cdot \nu }\left| \phi
\right\rangle  , \nu \in \mathbb{R}^{n}
\end{equation}

\noindent The unitary\ \ irreducible representations are labeled
by the $\nu _{i}$ that are the eigenvalues of the hermitian representation
of the basis $\{A_{i}\}$ of the abelian Lie algebra,
\begin{equation}
{\widehat{A}}_{i}\left| \phi \right\rangle  =\xi ^{\prime }( A_{i})
\left| \phi \right\rangle  =\nu _{i}\left| \phi \right\rangle  .
\end{equation}

\noindent \ \ The equivalence classes $[\xi ]\in {\text{\boldmath
$U$}}_{\mathcal{A}( n) }$ each have a single element $[\xi ]\simeq
\xi $ as, for the abelian group, the expression (131) is trivial.
The representations $\rho $ act on ${\text{\boldmath $\mathrm{H}$}}^{\xi
}\simeq \mathbb{C}$ and are one dimensional and therefore must\ \ commute
with the $\xi $. Therefore, in equation (130),\ \ $\rho ( g) \xi
( h) {\rho ( g) }^{-1}=\xi ( h) $ and (129) simplifies to
\begin{equation}
\xi ( a) =\xi (  \varsigma _{k}a) =\xi ( k a k^{-1}) , a\in \mathcal{A}(
m) ,\ \ k\in \mathcal{K}\mbox{}^{\circ}\text{}.\ \ %
\label{PH: Little group abelian equation}
\end{equation}
\begin{theorem}

{\bfseries {\upshape (Mackey).}} {\upshape Unitary irreducible representations
of a semidirect product with an abelian normal subgroup.}\ \ Suppose
that we have a semidirect product group {\upshape $\mathcal{G}\simeq
\mathcal{K}\otimes _{s}\mathcal{A}$} where $\mathcal{A}$ is abelian.
Let $\mathrm{\xi }$ be the unitary irreducible representation (that
are the characters) of $\mathrm{\mathcal{A}}$ on ${\text{\boldmath
$\mathrm{H}$}}^{\xi }\simeq \mathbb{C}$. Let $\mathcal{K}\mbox{}^{\circ}\subseteq
\mathcal{K}$ be a Little group defined by (134) with the corresponding
stabilizers\ \ $\mathcal{G}\mbox{}^{\circ}\simeq \mathcal{K}\mbox{}^{\circ}\otimes
_{s}\mathcal{A}$.\ \ Let $\mathrm{\sigma }$ be the unitary irreducible
representations of $\mathrm{\mathcal{K}}\mbox{}^{\circ}$ on the
Hilbert space ${\text{\boldmath $\mathrm{H}$}}^{\sigma }$. Define
the representation $ \varrho \mbox{}^{\circ}=\sigma \otimes \xi
$ of the stabilizer that acts on the Hilbert space\ \ ${\text{\boldmath
$\mathrm{H}$}}^{\varrho \mbox{}^{\circ}}\simeq {\text{\boldmath
$\mathrm{H}$}}^{\sigma }\otimes \mathbb{C}$.\ \ \ The theorem then
proceeds as in the case of the general Theorem 9.\label{PH: theorem:
Mackey abelian case}
\end{theorem}
\section{Appendix B: Polarized Realization of the Weyl-Heisenberg
group}

The maps $\varphi ^{\pm }$ defined in (35) is an isomorphism.\ \ Therefore
the $\Upsilon ^{\pm }( p,q,\iota ^{\pm }) $ are elements of the
Weyl-Heisenberg group realized in another coordinate system of matrices.
These realizations are referred to as the polarized realizations
\cite{folland}. The group products in these coordinates are computed
directly from\ \ (3-4) to be\ \ 
\begin{equation}
\begin{array}{l}
 \begin{array}{l}
 \Upsilon ^{+}( p^{\prime },q^{\prime },\iota ^{\prime }) \Upsilon
^{+}( p,q,\iota ) =\Upsilon ^{+}( p^{\prime }+p,q^{\prime }+q,\iota
+\iota +p^{\prime }\cdot q) , \\
 \Upsilon ^{-}( p^{\prime },q^{\prime },\iota ^{\prime }) \Upsilon
^{-}( p,q,\iota ) =\Upsilon ^{-}( p^{\prime }+p,q^{\prime }+q,\iota
+\iota -q^{\prime }\cdot p) , \\
 {\Upsilon ^{\pm }( p,q,\iota ) }^{-1}=\Upsilon ^{\pm }( -p,-q,-\iota
\pm p\cdot q) .
\end{array}
\end{array}
\end{equation}

\noindent Note that the polarized realizations factor directly
\begin{gather}
\Upsilon ^{+}( 0,q,\iota ) \Upsilon ^{+}( p,0,0) =\Upsilon ^{+}(
p,q,\iota ) ,
\\\Upsilon ^{-}( p,0,\iota ) \Upsilon ^{-}( 0,q,0) =\Upsilon ^{-}(
p,q,\iota ) .
\end{gather}

The existence of the isomorphisms $\varphi ^{\pm }$\ \ and these
two different normal subgroups $\mathcal{A}( n+1) $ with elements
$\Upsilon ( p,0,\iota ) $ and $\Upsilon ( 0,q,\iota ) $ whose intersection
is the center $\mathcal{Z}\simeq \mathcal{A}( 1) $ is responsible
for many of the remarkable properties of the Weyl-Heisenberg group.\ \ In
fact, we shall see shortly that the choice of the normal subgroup
in determining the unitary representations when applying the Mackey
theorems results in unitary representations with either $p$ or $q$
diagonal.\ \ 

\noindent The matrix realization corresponds to a coordinate system
of the Lie group and is therefore not unique.\ \ The polarized matrix
realizations are given by the $n+2$ dimensional square matrices
\begin{equation}
\Upsilon ^{+}( p,q,\iota ) =\left( \begin{array}{lll}
 1 & q^{\mathrm{t}} & \iota  \\
 0 & 1_{n} & p \\
 0 & 0 & 1
\end{array}\right) , \Upsilon ^{-}( p,q,\iota ) =\left( \begin{array}{lll}
 1 & p^{\mathrm{t}} & \iota  \\
 0 & 1_{n} & q \\
 0 & 0 & 1
\end{array}\right) .
\end{equation}
\section{Appendix C: Extended Central Extension}

The central extension for a group that is not connected group is
not necessarily unique.\ \ The central extension for a group that
is not connected may be defined by requiring exact sequences both
for the cover of the group and the homomorphisms onto the discrete
group for the components.\ \ \ For the\ \ $\mathbb{Z}_{2}\otimes
_{s}\mathcal{H}\overline{\mathcal{S}p}( 2n) $, these sequences are
\cite{Azcarraga}
\begin{equation}
\begin{array}{lllllllll}
 \ \  & \ \  & e &   & e &   & e & \ \  & \ \  \\
   &   & \downarrow  &   & \downarrow  &   & \downarrow  &   & 
\\
 e & \rightarrow  & \mathbb{Z}\otimes \mathcal{A}( 1)  & \rightarrow
& \mathcal{H}\overline{\mathcal{S}p}( 2n)  & \rightarrow  & \mathcal{I}\mathcal{S}p(
2n)  & \rightarrow  & e \\
   & \ \  & \downarrow  &   & \downarrow  & \ \  & \downarrow  &
&   \\
 e & \rightarrow  & \mathbb{D} & \rightarrow  & \mathbb{Z}_{2}\otimes
_{s}\mathcal{H}\overline{\mathcal{S}p}( 2n)  & \rightarrow  & \mathbb{Z}_{2}\otimes
_{s}\mathcal{I}\mathcal{S}p( 2n)  & \rightarrow  & e \\
   & \ \  & \downarrow  &   & \downarrow  &   & \downarrow  &  
& \ \  \\
   &   & \mathbb{Z}_{2} &   & \mathbb{Z}_{2} &   & \mathbb{Z}_{2}
&   &   \\
   &   & \downarrow  &   & \downarrow  &   & \downarrow  &   & 
\\
 \ \  &   & e &   & e &   & e & \ \  &  
\end{array}
\end{equation}

\noindent The solution is $\mathbb{D}\simeq \mathbb{Z}_{2}\otimes
\mathbb{Z}\otimes \mathcal{A}( 1) $.\ \ Therefore the central extension
of\ \ $\mathbb{Z}_{2}\otimes _{s}\mathcal{I}\mathcal{S}p( 2n) $
is unique and is given by\ \ $\mathbb{Z}_{2}\otimes _{s}\mathcal{H}\overline{\mathcal{S}p}(
2n) $.
\section{Appendix D: Homomorphisms}

Representations are homomorphisms of a group $\mathcal{G}$.\ \ If
the homomorphism is an isomorphism, then the representation is said
to be faithful and otherwise it is degenerate. Theorem 4 establishes
that degenerate representations are faithful representations of
groups homomorphic to $\mathcal{G}$. The homomorphisms can be characterized
by the normal subgroups that are the kernel of the homomorphism.

First we consider the subgroup $\mathcal{H}\overline{\mathcal{S}p}(
2n) $ that we have noted in (22) is the central extension of $\mathcal{I}\mathcal{S}p(
2n) $ with center 
\begin{equation}
\mathcal{Z}=\mathbb{Z}\otimes \mathcal{A}( 1) %
\label{MG: Center of aut}
\end{equation}

\noindent where $\mathbb{Z}$ is the center of $\overline{\mathcal{S}p}(
2n) $ and $\mathcal{A}( 1) $ is the center of $\mathcal{H}( n) $
(31).\ \ \ The double cover of $\mathcal{S}p( 2n) $ is the metaplectic
group $\mathcal{M}p( 2n) $. As $\mathbb{Z}_{2}$ is a normal subgroup
of $\mathbb{Z}$,\ \ that there is also a homomorphism from the cover
of the symplectic group to the metaplectic group \ \ 
\begin{equation}
\pi :\overline{\mathcal{S}p}( 2n) \rightarrow \mathcal{M}p( 2n)
,\ \ \ \ \ker ( \pi ) \simeq \mathbb{Z}/\mathbb{Z}_{2} .%
\label{MG: symplectic to metaplectic homomorphism}
\end{equation}

\noindent This gives the sequence of homomorphic groups where the
homomorphisms have kernels that are subgroups of the center $\mathcal{Z}$.\ \ \ \ 
\begin{equation}
\begin{array}{lllllll}
 \mathcal{H}\overline{\mathcal{S}p}( 2n)  & \rightarrow  & \mathcal{H}\mathcal{M}p(
2n)  & \rightarrow  & \mathcal{H}\mathcal{S}p( 2n)  &   &   \\
   & \searrow  &   & \searrow  &   & \searrow  &   \\
   &   & \mathcal{I}\overline{\mathcal{S}p}( 2n)  & \rightarrow
& \mathcal{I}\mathcal{M}p( 2n)  & \rightarrow  & \mathcal{I}\mathcal{S}p(
2n) .
\end{array}%
\label{MG: homomorphic central sequence for HSp}
\end{equation}

The group $\mathcal{I}\mathcal{S}p( 2n) $ that has a trivial center
terminates the sequence.\ \ It is the maximal {\itshape classical}
symmetry group.\ \ The projective representations of any of the
groups in this sequence is equivalent to the unitary representations
of the $\mathcal{H}\overline{\mathcal{S}p}( 2n) $.\ \ \ The above
expressions also apply to the full group $\mathbb{Z}_{2}\otimes
_{s}\mathcal{H}\overline{\mathcal{S}p}( 2n) $ by prefixing ``$\mathbb{Z}_{2}\otimes
_{s}$'' onto each of the groups that appear in (142).

In addition to the above homomorphisms that have abelian kernels,
we have the additional homomorphisms 
\begin{equation}
\pi :\mathbb{Z}_{2}\otimes _{s}\mathcal{H}\overline{\mathcal{S}p}(
2n) \rightarrow \mathcal{K}, \ker ( \pi ) =\mathcal{N},%
\label{MG: aut c to G}
\end{equation}

\noindent with
\begin{equation}
\begin{array}{ll}
 \mathcal{N} & \mathcal{K} \\
 \mathcal{H}( n)  & \mathbb{Z}_{2}\otimes \overline{\mathcal{S}p}(
2n)  \\
 \mathbb{Z}/\mathbb{Z}_{2}\otimes \mathcal{H}( n)  & \mathbb{Z}_{2}\otimes
\mathcal{M}p( 2n)  \\
 \mathbb{Z}\otimes \mathcal{H}( n)  & \mathbb{Z}_{2}\otimes \mathcal{S}p(
2n)  \\
 \mathcal{H}\mathcal{S}p( 2n)  & \mathbb{Z}\otimes \mathbb{Z}_{2}
\\
 \mathcal{H}\mathcal{M}p( 2n)  & \mathbb{Z}_{2}\otimes \mathbb{Z}_{2}
\\
 \mathcal{H}\overline{\mathcal{S}p}( 2n)  & \mathbb{Z}_{2}
\end{array}%
\label{MG: aut c homomorphisms}
\end{equation}\label{sd}\label{degH}\label{hermalg}\label{con}\label{sm}\label{sdirect}\label{pin}\label{sedirectd}


\begin{thebibliography}{000}
\bibitem{weyl} Weyl, H. (1927). \textit{Quantenmechanik und Gruppentheorie}.
Zeitscrift fur Physik, \textbf{46}, 1--46. \label{weyl}
\bibitem{folland} Folland, G. B. (1989). \textit{Harmonic Analysis
on Phase Space}. Princeton: Princeton University Press.\label{folland}
\bibitem{Stone} Stone, M. H. (1932). \textit{On one-parameter unitary
groups in Hilbert Space}. Annals Math. , \textbf{33}, 643--648.
\label{Stone}
\bibitem{vonNeumann} von Neumann, J. (1932). \textit{Ueber Einen
Satz Von Herrn M. H. Stone}. Annals. Math., \textbf{33}, 567--573.
\label{vonNeumann}
\bibitem{wigner} Wigner, E. P. (1939). \textit{On the unitary representations
of the inhomogeneous Lorentz group}. Annals of Math., \textbf{40},
149--204 . \label{wigner}
\bibitem{Weinberg1} Weinberg, S. (1995). \textit{The Quantum Theory
of Fields, Volume 1}. Cambridge: Cambridge.\label{Weinberg1}
\bibitem{bargmann} Bargmann, V. (1954). \textit{On Unitary Ray Representations
of Continuous Groups}. Annals Math., \textbf{59}, 1--46. \label{bargmann}
\bibitem{mackey2} Mackey, G. W. (1958). \textit{Unitary Representations
of Group Extensions. I}. Acta Math., \textbf{99}, 265--311. \label{mackey2}
\bibitem{degosson} de Gosson, M. (2006). \textit{Symplectic Geometry
and Quantum Mechanics}. Berlin: Birkhauser.\label{degosson}
\bibitem{Low13} Campoamor-Stursberg, R., \& Low, S. G. (2009). \textit{Virtual
copies of semisimple Lie algebras in enveloping algebras of semidirect
products and Casimir operators}. J. Phys. A, \textbf{42}, 065205.
\href{http://arxiv.org/abs/0810.4596}{http://arxiv.org/abs/0810.4596}
\label{Low13}
\bibitem{Major} Major, M. E. (1977). \textit{The quantum mechanical
representations of the anisotropic harmonic oscillator group}. J.
Math. Phys., \textbf{18}, 1938--1943. \label{Major}
\bibitem{mackey} Mackey, G. W. (1976). \textit{The theory of unitary
group representations}. Chicago: University of Chicago Press.\label{mackey}
\bibitem{weil} Weil, A. (1964). \textit{Sur certains groupes d'op\'erateurs
unitaires}. Acta {\itshape Math.}, \textbf{111}, 143--211. \label{weil}
\bibitem{Low5} Low, S. G. (2006). \textit{Reciprocal relativity
of noninertial frames and the quaplectic group}. Found. Phys., \textbf{36}(6),
1036--1069. \href{http://arxiv.org/abs/math-ph/0506031}{http://arxiv.org/abs/math-ph/0506031}
\label{Low5}
\bibitem{Low6} Low, S. G. (2007). \textit{Reciprocal relativity
of noninertial frames: quantum mechanics}. J. Phys A, \textbf{40},
3999--4016.  \href{http://arxiv.org/abs/math-ph/0606015}{http://arxiv.org/abs/math-ph/0606015}
\label{Low6}
\bibitem{Low14} Low, S. G. (2012). \textit{Relativity Implications
of the Quantum Phase}. J. Phys.: Conf. Ser., \textbf{343 }, 012069.
\label{Low14}
\bibitem{sternberg} Sternberg, S. (1994). \textit{Group theory and
physics}. Cambridge: Cambridge Press.\label{sternberg}
\bibitem{barut} Barut, A. O., \& Raczka, R. (1986). \textit{Theory
of Group Representations and Applications}. Singapore: World Scientific.\label{barut}
\bibitem{Azcarraga} Azcarraga, J. A., \& Izquierdo, J. M. (1998).
\textit{Lie Groups, Lie Algebras, Cohomology and Some Applications
in Physics}. Cambridge: Cambridge University Press.\label{Azcarraga}
\end{thebibliography}
\end{document}